\documentclass[12 pt preprint]{aastex}
\usepackage{rotating}
\usepackage[T1]{fontenc}
\usepackage[latin9]{inputenc}
\usepackage{array}
\usepackage{float}
\usepackage{graphicx}
\usepackage{amssymb}
\usepackage{subfigure}
\usepackage{multirow}
\usepackage{longtable}
\usepackage{longtable,lscape}
\makeatletter

\begin{document}

\title{A new method of pulse-wise spectral analysis of Gamma-ray Bursts}

\author{Rupal Basak and A.R. Rao}

\affil{Tata Institute of Fundamental Research, Mumbai - 400005, India. $rupalb@tifr.res.in,
arrao@tifr.res.in$}

\begin{abstract}

Time-resolved spectral analysis, though a very promising method to understand
the emission mechanism of gamma-ray bursts (GRBs), is difficult to implement 
in practice because of poor statistics. We present a new
method for pulse-wise time-resolved spectral study of the individual pulses of
GRBs, using the fact that many spectral parameters are either constants or smooth
functions of time. We use this method for the two pulses of GRB 081221, the brightest GRB
with separable pulses. We choose, from the literature, a set of possible models which includes the Band model, 
blackbody with a power-law (BBPL), a collection of black bodies with a smoothly 
varying temperature profile, along with a power-law (mBBPL), and two blackbodies 
with a power-law (2BBPL). First, we perform time-resolved study to confirm the 
spectral parameter variations, and then construct the new model to perform a joint
spectral fit. We find that any photospheric emission in terms
of black bodies is required mainly in the rising parts of the pulses and the falling
part can be adequately explained in terms of the Band model, with the low energy
photon index within the regime of synchrotron model. Interestingly,
we find that 2BBPL is comparable or sometimes even better, though marginally, than the Band model, in all episodes. 
Consistent results are also obtained for the brightest GRB of Fermi era --- GRB 090618. 
We point out that the method is generic enough to test any spectral model with well defined
parameter variations.

\end{abstract}


\section{INTRODUCTION}
The spectrum of the prompt emission of a gamma-ray burst (GRB) is generally fitted with the celebrated
Band spectral model (Band et al. 1993). This model represents a non-thermal spectrum, and it can be 
described in terms of two smoothly joined power laws.
Generally speaking, Band model adequately fits most of the time-integrated prompt emission spectra of GRBs,
though additional spectral components show up for some  GRBs (e.g., Preece et al. 1996; Gonzalez et al. 2003; Shirasaki et al. 2008).
But, to emphasise, the exceptions are very few in number in comparison with the large set of 
GRBs, which can be fitted with a simple Band only function (e.g., Kaneko et al. 2006; Nava et al. 2011).
Zhang et al. (2011) have found that 15 out of 17 LAT detected GRBs could be fitted with Band only 
model (category I, in their notation). 
Hence, the Band function, 
till date, is the simplest, standalone model for GRB spectra, 
whether time-integrated or time-resolved.

Although the Band model is statistically the most appropriate model of GRB data, its physical origin is yet to be identified. 
Over the years, many authors have investigated the underlying mechanism of the prompt emission.
In the fireball model of GRB, the observed radiation during the prompt emission is attributed 
to a highly relativistic optically thick outflow, which thermalizes photons due 
to random collisions. This thermal energy is expected to be seen from the photosphere (Goodman 1986; Paczynski 1986;
Meszaros \& Rees 2000; Pe'er 2008), where the fireball becomes optically thin and radiation decouples from the matter. 
In this scenario, the light curve (LC) is expected to be a single, smooth pulse and the spectrum should be a Lorentz 
boosted BB, while the temperature of the BB adiabatically cools. Hence, the time-integrated spectrum should 
be a superposition of many BBs. Though some GRBs do have a single pulse, most GRB light curves are either superposition of many pulses or 
they are highly variable. Moreover, A BB in the Rayleigh-Jeans region has a photon index 1.0, which is much harder than that actually 
observed for GRBs ($\alpha \sim -1$). 
Many models have been proposed to overcome these difficulties. For example, the internal shock model 
(Rees \& Meszaros 1994; Woods \& Loeb 1995; Sari \& Piran 1997; Kobayashi et al. 1997) assumes that the major radiation is not
due to photospheric emission, but optically thin synchrotron radiation (SR) from internal shocks. One major problem with the 
synchrotron radiation is that the low energy photon index is limited to $\alpha < -\frac{2}{3}$ (Preece et al. 1998). Crider, 
Liang \& Preece (1998), using time-resolved spectra of 99 GRBs, have shown that the instantaneous spectra and their evolution 
cannot be explained by SR --- $\alpha$ often crosses the line of death, set by the synchrotron model. 
Another possibility is that the radiation is due to inverse Compton (IC) of the thermal photons near the photosphere 
(Thompson 1994; Pe'er et al. 2005, 2006; Beloborodov 2010; Lazzati \& Begelman 2009). This process can indeed produce a Band 
like spectrum, but with a rather hard value of $\alpha \sim 0.4$ at the best, assuming slow heating. 

A unique prompt emission model of a GRB is yet to be settled. From the phenomenological point of view,
the correct model can be identified by segregating the details from the average properties.
For example, GRBs are superpositions of pulses (Nemiroff 2000; Norris 2005). Hence, one should use the individual 
pulses for spectral study, instead of the full GRB. The next step is to study the spectral evolution within the individual pulses.
Hence, one should do time-resolved spectroscopy in order to extract greater information than merely
an average spectral property of a pulse, e.g., average peak energy, isotropic energy etc. But, performing such a detailed study is 
difficult, as one loses photon counts. For example, Ghirlanda et al. (2010) have done time-resolved spectral study of  9 selected 
GRBs, detected by Gamma-ray Burst Monitor (GBM) onboard the Fermi satellite.
As the photon count is low in each time bin, they could model the spectra only with a cut-off powerlaw. Parameters of a more complicated model,
such as Band, cannot be well constrained from the time-resolved data. The solution to this problem lies in the realisation that
spectral evolution is not totally unpredictable, and one can suitably parametrize this evolution in order to reduce the number
of free parameters of the description. For example, the spectral evolution of a GRB pulse is generally described as a hard-to-soft evolution
(e.g., Liang \& Kargatis 1996; Kocevski \& Liang 2003; Nemiroff 2012). Recently, Basak and Rao (2012a; b) have assumed 
this hard-to-soft evolution of the individual pulses of the set of 9 GRBs of Ghirlanda et al. (2010) to generate simultaneous 
spectral and timing model of the pulses, with essentially two parameters, namely, the peak energy at the start of the pulse 
($E_{peak,0}$) and the characteristic evolution parameter ($\phi_0$). 
The basic assumptions in this approach, however, are not well established. For example, it was assumed that
the spectral softening happens throughout the pulse, though there are evidences that some GRB pulses 
show a different behaviour  like the intensity tracking spectral evolution (see e.g., Lu et al. 2012). 
Further, it was also assumed that the applicable model is the Band function throughout. Hence, it is essential
to critically examine all the applicable spectral models and their evolution to 
arrive at a correct pulse-wise description of a GRB.

In this paper, we discuss a new method for pulse-wise spectral analysis where we 
parametrize the spectral evolution in order to arrive at the correct spectral description with a minimum set of free parameters. 
We apply this method 
to study GRB 081221, the brightest GRB with  clean, separable pulses. 
We compare the results obtained for this GRB with those for GRB 090618 --  the brightest GRB in the Fermi era.
%
The organisation of this paper is as follows ---
in Section 2, we describe the analysis techniques and the basic assumptions of our model. Results are given in  Section 3, 
and major conclusions are drawn in Section 4. 

\section{ANALYSIS METHOD}

\subsection{Data selection and  analysis}
The basic necessity for a good spectral analysis of GRB pulses is wide band coverage to identify
additional spectral components. The Gamma-ray Burst Monitor (GBM)  on board the Fermi satellite, with its wide
band width and excellent sensitivity, provides a good data base for such studies. 
It has two scintillation detectors: the sodium iodide (NaI) detector is sensitive in the
$\gtrsim$ 8 keV to $\sim$ 900 keV range while the BGO energy range is $\sim$ 200 keV to $\sim$ 40 MeV
(Meegan et al. 2009). 

\begin{figure}\centering
{

\includegraphics[width=6.5in]{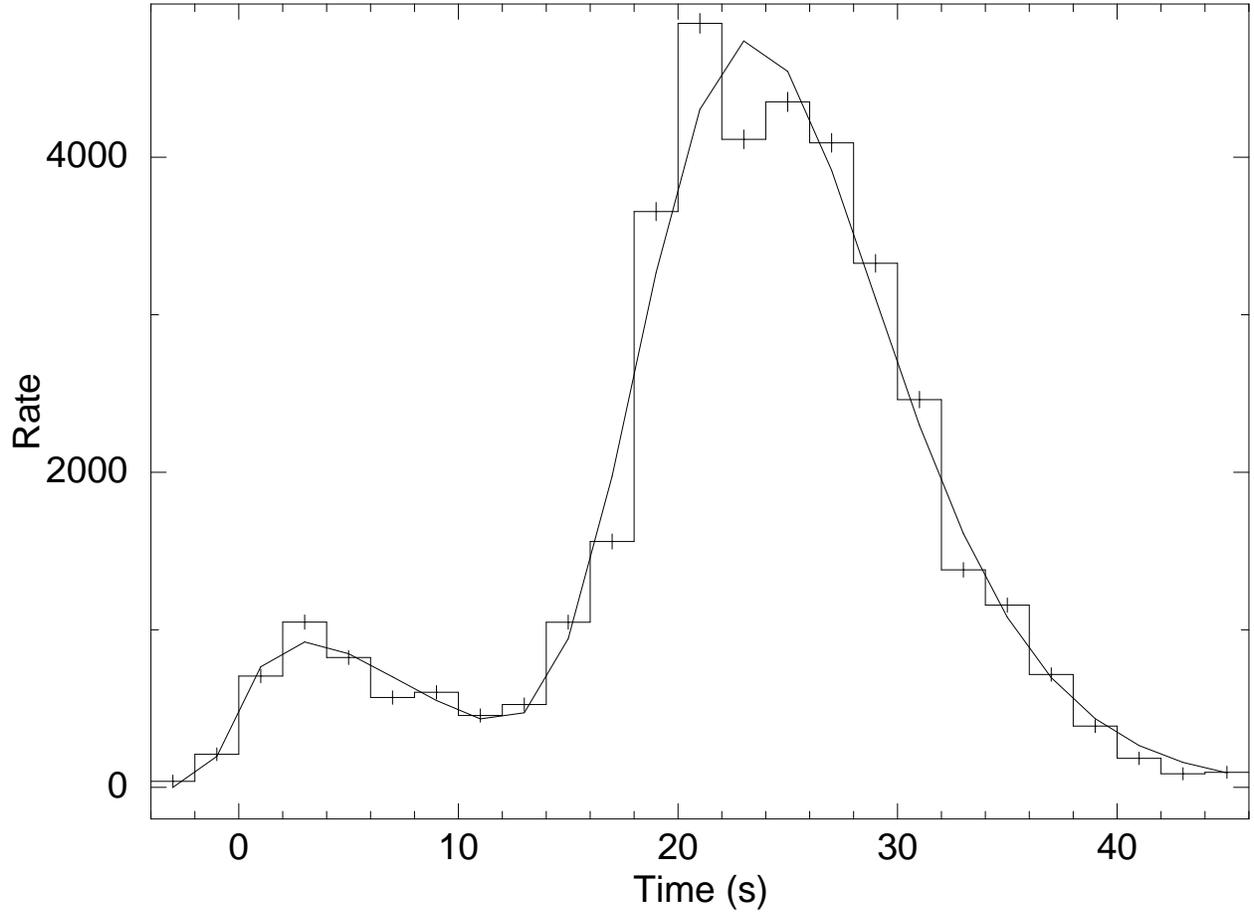} 

}
\caption{Background subtracted Light curve (LC) of GRB 081221, fitted with Norris model
(Norris et al. 2005). The LC is generated by adding the two highest count NaI (n1 and n2)
and one BGO (b0) detectors after binning by 2 s.
}
\label{LC}
\end{figure}

We examined the  Nava catalog (Nava 
et al. 2011) of Fermi/ GBM GRBs and found that there are 112 bright (fluence $\geq 10^{-6}$ erg), long ($\delta t \geq 15$ s) GRBs
and 11 of these GRBs have single/ separable pulses. GRB 081221 is the brightest among them. In Figure~\ref{LC}, we have 
plotted the light curve (LC) of this GRB with Norris model (Norris et al. 2005) fitted for the two pulses. 
We have also made a systematic analysis of the other 10 GRBs and the time-integrated spectral
analysis for all of them is given later.

We use the CSPEC data for time-integrated study and the time tag event (TTE) data 
for the time-resolved spectral analysis. We choose 2 or more NaI detectors having high count rate and one/ both BGO detector (s). 
For source selection and background subtraction, we use the \textit{rmfit v3.3pr7} tool, developed by User Contributions of Fermi 
Science Support Center (FSSC). The background exposure time is chosen before and after the burst. This background is modelled by
a polynomial of different degrees, according to the need. The PHA files are binned in energy channels so as to get a minimum count in 
each spectral bin. Typically, the NaI detectors are binned by minimum count $\gtrsim 40$, while the BGO detectors are binned by requiring
a minimum count of $\sim 50-60$. First, we perform time-integrated analysis for all the 11 bright, long 
GRBs having single/ separable pulses. We implement both C-stat and $\chi^2$ minimization methods in \textit{rmfit}. We fit either 
Band or Cut-off powerlaw (CPL) model. 

In Table 1, we report the best fit parameter values along with the corresponding 3$\sigma$ errors.
The reduced C-stat and $\chi^2_{red}$ with degrees of freedom (dof) are also reported.
For comparison, we quote the results of Nava et al. (2011) for these GRBs. It is clearly seen that 
all these values are matching quite well. The sources of very minor deviations between the values of this work, 
done by C-stat minimisation and Nava et al. (2011) are: (i) mismatch between actual start and stop time,
(ii) exact background selection and modelling and, (iii) the exact number of detectors and the channels used.
Comparing the deviation of the parameter values, it is clear that
the deviation due to using different statistics other than C-stat, i.e. $\chi^2$ minimisation, is much less than that due to
these other reasons. Hence, we conclude that the statistics plays a minimal role in actual parameter estimation.
In fact, the GRBs taken in our analysis are all bright GRBs (fluence $\geq 10^{-6}$ erg). Hence, by default, the
$\chi^2$ minimisation is a correct technique for parameter estimation of GRBs with high count rates.

\subsection{Spectral models for time-resolved study}

We select four models for the spectral study: a) Band model, b) blackbody with a power-law, c)
a modified blackbody with a power-law and d) two blackbodies with a power-law.
The Band function can be written in terms of the spectral  indices ($\alpha$ and $\beta$)
and the peak energy ($\rm E_{peak}$) as:

\begin{equation}
I(E) = \left\{ \begin{array}{ll}
 A_{b}\left[\frac{E}{100}\right]^{\alpha}exp\left[\frac{-(2+\alpha)E}{E_{peak}}\right] ~~~~~~~~$\rm ;~if$ ~~E\le \frac{(\alpha-\beta)}{(2+\alpha)}E_{peak}\\
\\  
A_{b}\left[\frac{E}{100}\right]^{\beta}exp\left[\beta-\alpha\right]\left[\frac{\left(\alpha-\beta\right)E_{peak}}
     {100\left(2+\alpha\right)}\right]^{(\alpha-\beta)} \rm ;~otherwise
       \end{array} \right.
\label{Band}
\end{equation}

Here, $A_b$ is the normalisation constant. 
A model consisting of 
thermal (blackbody with temperature kT) and non-thermal (power law with index $\Gamma$) 
components has been used earlier (see, e.g., Ryde 2004, Ryde et al. 2006, Ryde and Pe'er 2009, Pe'er and Ryde 2011).
We name this function as BBPL. This can be written as:
\begin{equation}
I(E)=\frac{K_{1}\times8.0525E^{2}}{(kT)^{4}\left[exp\left(E/kT\right)-1\right]}+K_{2}E^{-\Gamma}
\label{BBPL}
\end{equation}
where K1 and K2 are normalisation constants.
There are suggestions (e.g., Ryde et al. 2010) in the literature of a modified blackbody (mBB), which may exist due to angular dependence 
of the optical depth and the observed temperature (Pe'er 2008). Hence, we also investigate this model with a powerlaw (mBBPL). 
The mBB model is a multi-colour blackbody disk model; the local disk temperature kT(r) is proportional to $r^{-p}$.
In several GRBs there are distinct additional thermal components (see eg. Shirasaki et al. 2008)
and further,
if the GRB spectrum is due to thermal inverse Compton (IC) of seed photons, then there may be, in principle, multiple 
photon baths. For brevity, we take one more BB component and call it 2BBPL model. This has essentially two BBs with two
temperatures --- $kT_h$ and $kT_l$ and two normalisations --- $K_{1h}$ and $K_{1l}$. We use all the four models for our
subsequent analysis.
We note here that it is also possible to have combinations of the above models (like Band+PL,
Band + BB + PL), but, at this juncture, we consider only these four generic models. This 
is essentially because of the fact that the time resolved spectra of GRBs generally
consists of a broad peak with wings, which can be adequately captured by any of the above four models.

\subsection{Assumptions of the new pulse-wise spectral study}
Time-resolved spectral studies require a large number of parameters. If `n' is the number of time bins, then a four-parameter model, 
such as Band or BBPL, requires 4n parameters for a full description. Our motivation, in this study, is to reduce the number of parameters 
with some reasonable assumptions. The basic assumption we make is that the temperature (kT) and
the peak energy ($E_{peak}$) follow smooth time evolution. This evolution has 
a break at the peak of the pulse (see also Ryde and Pe'er 2009). In the following discussion we shall refer to them as two episodes --- 
rising and falling part. We assume that the time evolution law of kT and $E_{peak}$ is simple powerlaw of time 
i.e., $\sim t^{\mu}$, with different $\mu$ in different episodes. If `m' and `n' are the number
of time bins in these two episodes, respectively, then this parametrization reduces the number of free parameters by `m+n-2'.
Note that, ideally one should use one more parameter to account for the start time, but
we have chosen it to be zero for all the cases, except for the rising part of GRB 090618, where the pulse starts from -1 sec.
Hence, only for this case, we have chosen this start time to be -10 sec. As the parametrization is not corrected for the
actual start time, the actual values of $\mu$ are not unique and hence, these values should not be used to compare
different pulses. We also assume that for the  Band model the photon indices ($\alpha$ and $\beta$) are constants in each episode. 
Their values can be determined by simultaneously fitting all the time-resolved spectra in an episode, with the photon indices tied. 
For BBPL model, we assume that the index ($\Gamma$) of the powerlaw (PL) is constant in each episode. This index can be determined 
in the similar way as the photon indices of the Band. The parametrization of the norm of BBPL model is more complicated, as, unlike 
Band, we have two norms. In the Band model, norm is a free parameter. In order to have equal number of parameters as Band, either we 
can use an overall free norm with suitable parametrization of BBPL norms, or, we can parametrize one of the norms and treat the other
as a free parameter. Ryde and Pe'er (2009) have shown that the parameter 
${\cal{R}} = (F_{BB}/\sigma T^4)^{1/2}$ either increases with time or remains constant. If we assume that the observed flux of 
the BB varies as a simple function of time as $F_{BB} \sim t^{\zeta}$, then the variation of ${\cal{R}}$ can be parametrized as 
$\sim t^{\zeta/2-2\mu}$. For $\mu \le \zeta/4$, ${\cal{R}}$ will show the expected behaviors. The BB flux variation can be translated 
to the norm variation of Equation~\ref{BBPL} as $K_1 \sim t^{\nu_1}$. Now, if we parametrize the BB norm and treat the powerlaw (PL) 
norm as a free parameter of the model, we have the same number of free parameters as Band model. But, this makes the BBPL more 
constrained, in the sense that the overall norm is not a free parameter, as in the case of Band model. To overcome this difficulty, 
let us assume (which will be justified later) that the powerlaw (PL) component has norm variation as smooth as that of BB, i.e., 
$K_2 \sim t^{\nu_2}$. Then the overall norm (K) can be made a free parameter by parametrizing the ratio of the BB norm and PL norm as 
$K_1/K_2 \sim t^{\frac{\nu_1}{\nu_2}} \sim t^{\nu}$.

Thus, both Band and BBPL models have equal number of free parameters 
--- `m+n+8'. For Band model, `m+n' are norms of Band function in `m+n' time bins. The other 8 parameters are $\alpha$, $\beta$, $\mu$
and the $E_{peak}$ at the starting bin, $E_{peak}(t_0)$, in the two episodes. Peak energy at any time `t' is determined by 
$E_{peak}(t_0)\times(t/t_0)^{\mu}$. For the BBPL model, `m+n' are overall norms (K) and the other four
parameters are powerlaw index ($\Gamma$), $\mu$, $\nu$ and kT at the initial bin, kT($t_0$), in the two episodes. For mBBPL model, 
we have an added parameter, namely, p, in each episode. Hence, the number of parameters is `m+n+10'. For 2BBPL model, the photons are 
boosted by the same material, hence, the BB parameters cannot be arbitrary. We assume that the ratio of temperatures and norms of these 
BBs are fixed, which should be determined by tying the ratios in all bins. Hence, the number of parameters in this model is `m+n+12'. 
As an example, if there are 5 time bins preceding to the peak of a pulse and 10 bins afterwards, then the parametrization for Band and 
BBPL reduces the number of free parameters from 60 to 23; for a total of 25 bins this factor is 100 to 33 and so on.

To compare the significance of a model with respect to another, we have performed the F-test. The F value is defined in most general
case as, $F=\frac{\chi^2_1/dof_1}{\chi^2_2/dof_2}$, where index 1 is used for the primary model, while 2 is used for the alternative model.
We compute the probably (p) of a given F value and thereby find the $\sigma$ significance (and the \% confidence level) of the alternative
model preferred over the primary model. If the primary model is a subset of the alternative model, then the F value is defined as,
$F= \frac{(\chi^2_1-\chi^2_2)/(dof_1-dof_2)}{\chi^2_2/dof_2}$ and the difference between the dofs (M) is deemed as the dof of the
primary model.

\section{RESULTS}

\subsection{Time-resolved spectra of GRB 081221}

We start with a time-resolved spectral analysis so that some of the assumptions sketched above can be
examined and validated.
The major challenge in time-resolved spectroscopy is to define the time bin size, which crucially depends on two factors --- a) timescale
of spectral evolution and b) the minimum bin size allowed by the data, which in turn depends on the 
subjective decision of signal-to-noise ratio (SNR). One cannot violate the latter condition for a given SNR. 
The peak count rate of this GRB is $\sim$ 4000, while we have demanded that each 
PHA bin should have at least 40 counts (SNR $\sim$ 6.3, i.e., total $\sim$ 5000 counts, for 128 channels). Hence, we cannot choose 
smaller than $\sim$ 1 sec time bin. First, we choose uniform time bins of 3 second to extract time-resolved spectra. Later, we reduce 
the time bin to 1 second to check any improvement due to finer time bins. 

\subsubsection{Case I: Bin size of 3.0 seconds}
 In Table~\ref{081221_a}, we report the results of Band and BBPL fit to the time-resolved data
of 3 second bin size. The time bin starts from -1 sec, with 14 bins (numbered 0 to 13). Approximately, first 4 bins belong
to pulse 1, last 8 bins belong to pulse 2 and the 2 intermediate bins belong to the overlapping region. For the BBPL fit, we first fit 
the spectra with the powerlaw index ($\Gamma$) free. We note that $\Gamma$ is more or less constant for the major portion of the burst 
(bin \#0-2 for pulse 1 and \# 6-11 for pulse 2). We take the average of $\Gamma$ over these bins, separately for the two pulses, and 
found in both cases, $\Gamma$=1.83 with standard deviation ($\sigma$) 0.14 and 0.10 for pulse 1 and 2, respectively. $\Gamma$ has 
large error bars in the last bin of pulse 1 and last 2 bins of the second pulse. The values in the overlapping region (\#4 and \#5), 
apart from having large errors, may be ambiguous, and hence neglected. We freeze $\Gamma$ to 1.83 for all the bins and redo the analysis. 
The corresponding values are also reported in Table~\ref{081221_a}. Note that, by doing this we are gaining one degree of freedom in each 
time bin. It is clear from Table~\ref{081221_a} that for the second pulse, the time-resolved spectral fit with the Band model are much 
better than those of the BBPL fit in terms of the $\chi_{red}^{2}$. On the contrary, these values are comparable in the first pulse. 
Hence, the first pulse may be dominated by thermal emission. Note, however, that $\chi_{red}^{2}$ of BBPL is poor in bin \#1 of the 
first pulse, which has, in fact, the highest flux. Hence, this points to a different radiation mechanism than a simple BB, which may appear in 
the high flux regions.

\begin{figure}\centering
{

\includegraphics[width=6.5in]{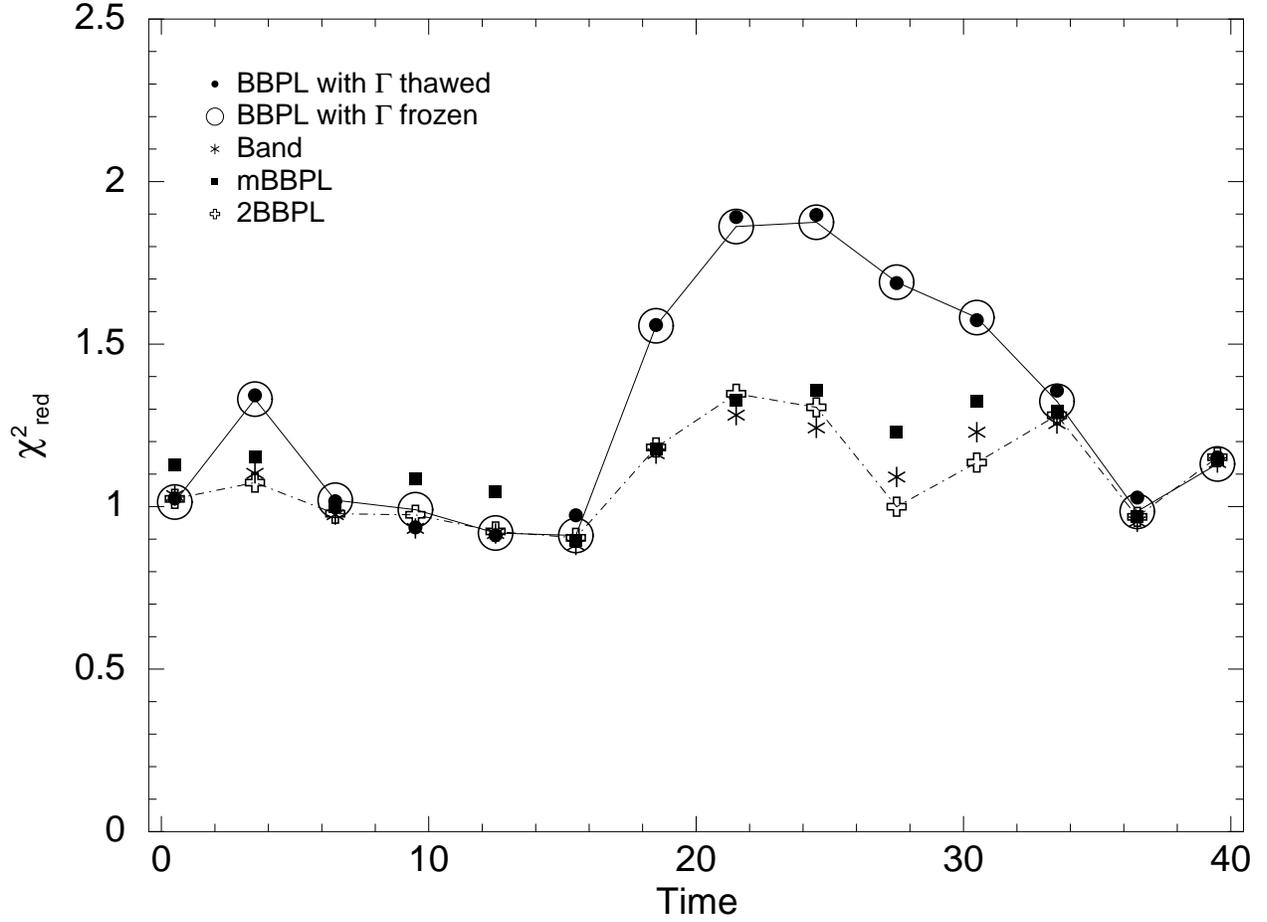} 

}
\caption{Comparison of $\chi^2_{red}$ of Band, BBPL, mBBPL and 2BBPL in the time-resolved spectroscopy of GRB 081221.
The symbols are explained in the inset. For convenience, we draw line to join the BBPL ($\Gamma$ free) and dot-dashed
line to join 2BBPL points. 
}
\label{chi}
\end{figure}

In Figure~\ref{chi}, we have plotted the $\chi_{red}^{2}$ of BBPL with $\Gamma$ thawed, BBPL with $\Gamma$ frozen
and Band fit with filled circles, open circles and stars, respectively. The BBPL model is clearly inferior to the Band model for a major portion of 
the second pulse, specially in the regions where photon counts are high. In the first pulse (-1 to 11 sec), the BBPL model, except for
the one bin is comparable to the Band model. We also fit mBBPL and 2BBPL, which are shown by filled boxes and pluses. The mBBPL and 2BBPL are 
as good as the Band model, in terms of $\chi_{red}^{2}$. Hence, the correct model for the major portion of the burst is either of these
three. 

\subsubsection{Case II: bin of size 1.0 second}
In the time-resolved analysis of Section 3.1.1, the Band and other models show superior fits compared 
to the BBPL model in terms of 
$\chi_{red}^{2}$. But, as Zhang (2011) pointed out for GRB 090902B, this can be an effect of evolution of the BBPL within a time bin. 
To investigate this, we make finer time bins of 1 second and redo the time-resolved spectroscopy. 
Note that this is the finest bin possible for a SNR $\sim$ 6.3. In Table~\ref{av_chi}, 
we report the average values of $\chi_{red}^{2}$ for different model fittings, obtained for different bin sizes. Apparently, there are 
improvements in the $\chi_{red}^{2}$ for finer bins, but we also see that these improvements are of the same orders for different models. 
Hence, it seems that the spectrum is not due to the evolution of BBPL, but one of these other models. But, 
all of these models are comparable. Hence, we cannot hope to find the correct model by taking finer bins.
Presence of a simple blackbody (BB) is apparent in the first pulse, with the exception of the second bin, where one
of the other models is correct. The second pulse is dominated either by a modified blackbody or a fully non-thermal 
radiation (Band) or inverse Compton (2BBPL). In order to find the right answer, one should carefully parametrize the
spectral evolution in each episode of the pulses separately and reduce the set of free parameters in the description.

\subsection{Parameter evolution}
\begin{figure}\centering
{

\includegraphics[width=6.5in]{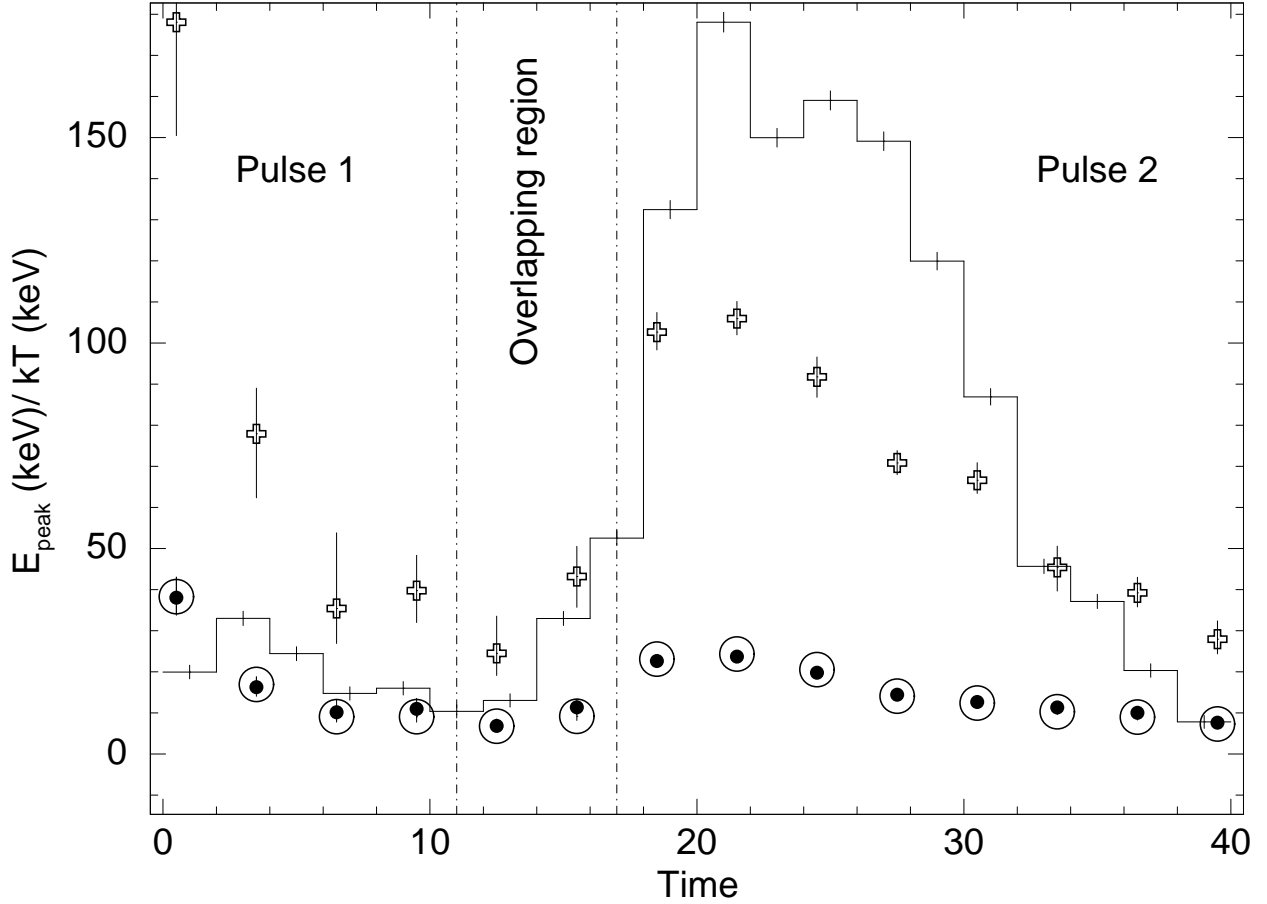} 

}
\caption{Time evolution of kT and $E_{peak}$ of GRB 081221. The filled circles are the BBPL with $\Gamma$ free, while the
open circles are BBPL with $\Gamma$ frozen. The pluses are $E_{peak}$. The light curve (LC) of the GRB is
shown as a histogram with errors. We have divided the LC into three regions. We see a clear hard-to-soft evolution in pulse 1. The same evolution is seen
in the falling part of the second pulse. In the overlapping region, the variation is rather soft-to-hard. This might be the 
effect of overlapping, or the evolution may as well be genuinely intensity tracking (see text).
}
\label{t_kt_epeak}
\end{figure}

The spectral evolution in the pulses are not arbitrary. For example,
the temperature of the blackbody evolves with time in a smooth way. This can be seen from Figure~\ref{t_kt_epeak}, where we
have plotted the kT of BBPL (both $\Gamma$ free and frozen) by circles. The peak energy variation is shown by pluses. The
parameters evolve smoothly as a function of time. For a single pulse, Ryde and Pe'er (2009) have shown that the temperature remains
constant or slowly declines with an average powerlaw index, $\langle a_T \rangle=-0.07$ with $\sigma(a_T)=0.19$, during the 
rise of the pulse and decays faster with an average index $\langle b_T \rangle=-0.68$ with $\sigma(b_T)=0.24$. The break 
time of this evolution has a strong positive correlation with the pulse peak time. We find a similar behaviour for both kT and 
$E_{peak}$ evolution. Hence, the spectral evolution can be described by a simple time evolution of temperature ($kT \sim t^\mu$)
or the peak energy ($E_{peak} \sim t^\mu$). The index, $\mu$, in principle may have two values in the two episodes, namely,
the rising and the falling part.

\begin{figure}\centering
{

\includegraphics[width=6.5in]{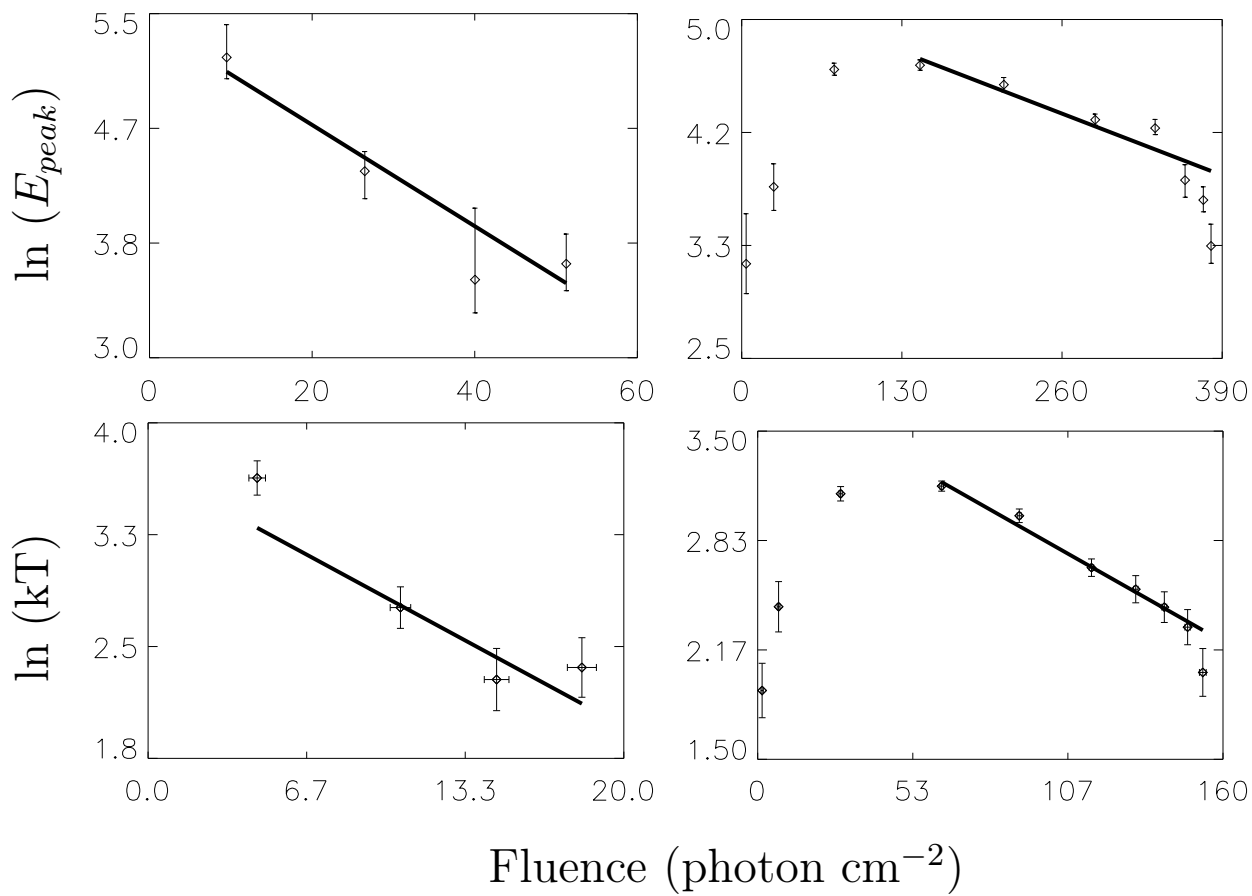} 

}
\caption{Verification of Liang \& Kargatis (1996, LK96) law in the pulses of GRB 081221. The first pulse (left 
panels) shows a strict hard-to-soft evolution,
while the second pulse is intensity tracking. The x-axis represents the ``running fluence'', defined by LK96. 
For the BBPL model, fluence here means that of the BB component. }
\label{kt_epeak_fluence}
\end{figure}

Liang and Kargatis (1996; hereafter LK96) showed for FRED pulses that the peak energy of the EF(E) (or $\nu F_{\nu}$) spectrum
follows a more complicated evolution. $E_{peak}$ decreases exponentially with the running fluence as:

\begin{equation}
 E_{peak}(t)=E_{peak,0}~exp\left(-\frac{\phi_{Band}(t)}{\phi_{Band,0}}\right)
\label{LK96_law}
\end{equation}

where, $E_{peak,0}$ and $\phi_{Band,0}$ are the constants of the $\rm E_{peak}$ evolution law,
$\phi_{Band}(t)$=$\int_{t_0}^t f(t')dt'$, is the fluence at time t, $f(t')$ being the flux.
In Figure~\ref{kt_epeak_fluence} (upper panels), we have plotted the ln ($E_{peak}$) with the fluence. Note that the fluence of each pulse is
calculated from their respective start time. Clearly, the $E_{peak}$ evolution strictly follows the LK96 law in the first pulse.
In the second pulse, however, the variation is not smooth throughout. In the falling part, the variation is clearly LK96 type, but
in the rising part, the variation is rather ``soft-to-hard''. This effect may come due to overlap between two pulses. 
In fact, the first two bins of the second pulse belong to the overlapping region. Hence, they might be contaminated with the preceding pulse. 
However, the third bin, where there should not be any effect of the first pulse, also deviates from the LK96 law. This might indicate that the 
second pulse is genuinely intensity tracking. Kocevski and Liang (2003)
argued that the ``intensity tracking'' pulses for which this evolution does not appear very prominent,
are rather made of more than one short hard-to-soft pulses. Ghirlanda et al. (2011), on the other hand, have analysed 
time-resolved spectra of 11 long and 12 short Fermi GRBs and found that the long GRBs appear 
to follow a ``soft-hard-soft'' trend, tracking the flux of the GRB, rather than a strict ``hard-to-soft'' evolution. Lu et al. (2012)
have categorised GRB 081221 as one having a strict ``hard-to-soft'' pulse followed by ``intensity-tracking'' pulse. They have simulated
overlapping pulses to show that in the overlapping region, the spectral evolution may appear ``intensity-tracking''. However, they also
found some single pulses to have ``intensity-tracking'' spectral evolution. Hence, the second pulse may be genuinely ``intensity tracking''.

In Figure~\ref{kt_epeak_fluence}  bottom panels, the temperature evolution is plotted against blackbody fluence. Here the same behaviour 
is noticed. Hence, kT
evolution of the first pulse and the falling part of the second pulse can as well be described by a similar exponential decay:

\begin{equation}
kT(t)=kT_0~exp\left(-\frac{\phi_{BB}(t)}{\phi_{BB,0}}\right)
\label{LK_new}
\end{equation}

where $\phi_{BB}(t)$=$\int_{t_0}^t f(t')dt'$ is the running fluence of the blackbody component at time t, $f(t')$ being the flux at $t'$. 
$\phi_{BB,0}$ and $kT_0$ are the constants of the evolution law. 
LK96 law is empirical and a simpler version, in principle, can be used instead, such as a simple powerlaw of time. As fluence is a 
monotonically increasing function of time, either of them can be used for evolution study.

\begin{figure}\centering
{

\includegraphics[width=6.5in]{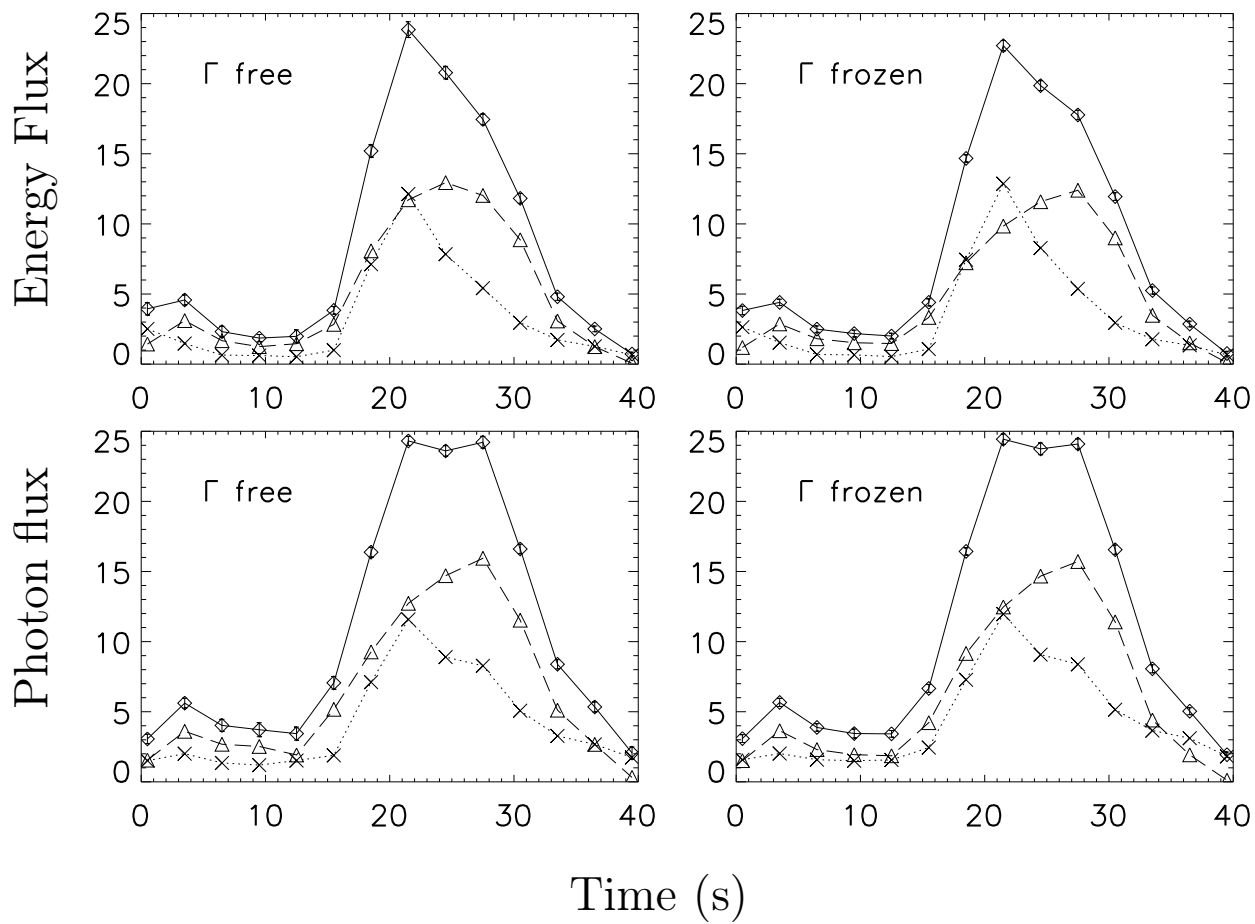} 

}
\caption{Flux evolution of GRB 081221. Crosses represent BB flux, while triangles represent PL flux. The total flux is marked by open
boxes. Energy flux, in the units of $10^{-7}$ erg cm$^{-2}$ s$^{-1}$, is plotted in the upper panels; photon flux, in the units of
photon cm$^{-2}$ s$^{-1}$, is plotted in the lower panel, for both $\Gamma$ free and frozen cases.
}
\label{flux_evolution}
\end{figure}

\begin{figure}\centering
{

\includegraphics[width=6.5in]{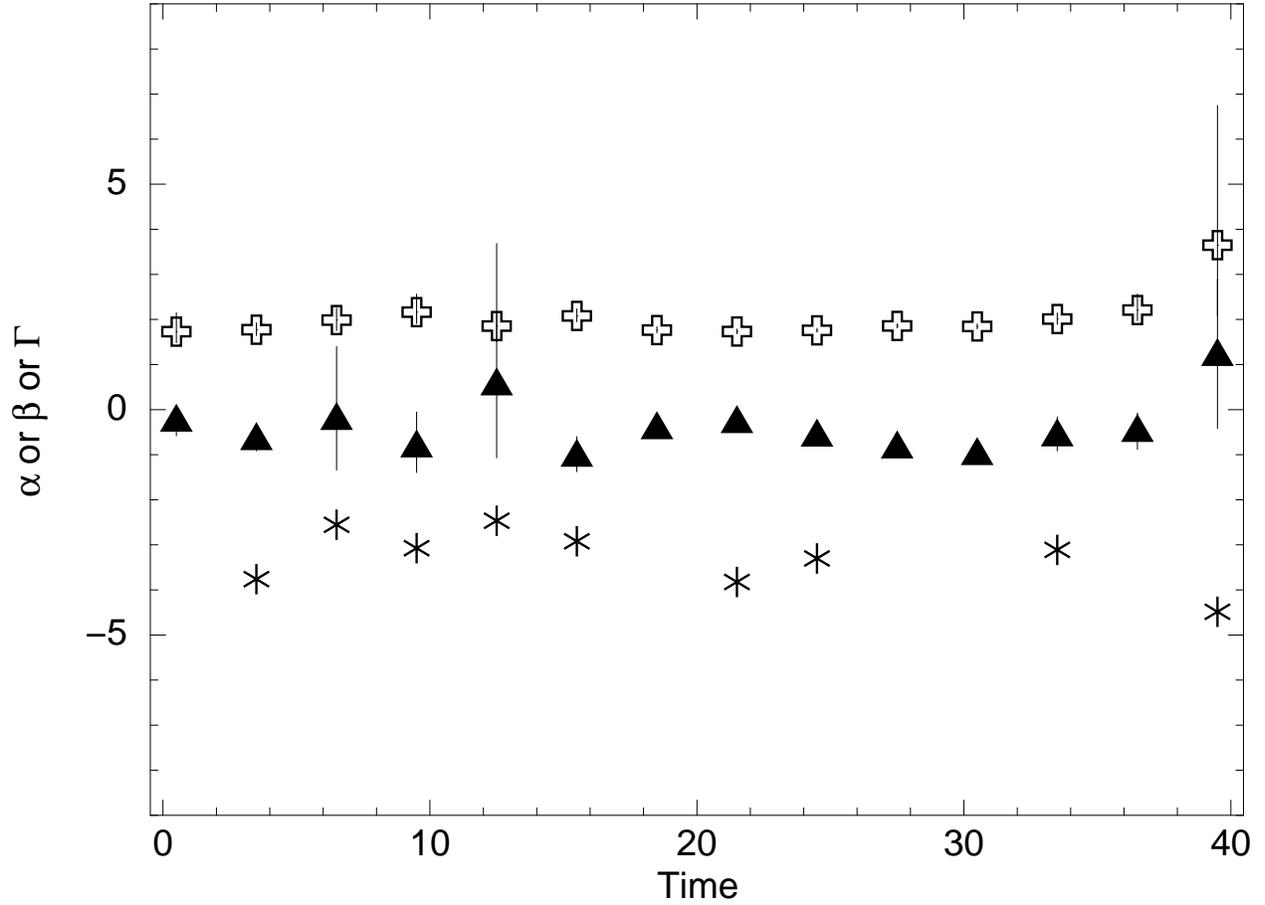} 

}
\caption{Evolution of $\alpha$ (triangles), $\beta$ (stars) of Band and $\Gamma$ (pluses) of BBPL throughout the GRB. The error bars of 
parameter $\beta$, being large and undetermined in many cases, are not shown. In some cases, the $\beta$ pegs at -10.0. Notice that 
the parameters are more or less constant throughout the burst. Hence, they can be tied (see text).
}
\label{par_evolution}
\end{figure}

The BBPL model has two components. Hence, in order to parametrize the norms of this model one has to see the flux evolution of
the individual components. In Figure~\ref{flux_evolution}, we have plotted both photon and energy flux of the individual components calculated
for 8-900 keV energy range. The flux evolutions look similar for both $\Gamma$ free and frozen cases. Interestingly, the PL flux 
is as smooth as the BB flux, in each pulse. Hence, as argued in Section 2.3, we can safely assume that the ratio of their evolutions
is a smooth function of time. In Figure~\ref{par_evolution}, we have shown the evolution of $\alpha$, $\beta$ of Band and $\Gamma$ of BBPL.
The parameter $\beta$, in many cases, have either large error bars or only an upper limit could be derived.
In some cases, they peg to the value -10. It is clear from Figure~\ref{par_evolution} that the parameters remain reasonably constant
at all episodes (rising and falling part) of a pulse. Hence, we tie them over all the time bins in a given episode to determine them with
greater accuracy. This reduces the number of free parameters of the description of spectral evolution to a great extent, as 
described in Section 2.3.

\subsection{Results of the parametrized spectral fitting of GRB 081221}
The fact that the parameters are well behaved functions of time, makes the time-resolved spectroscopy more tractable, as we can
reduce the number of free parameters in the pulse-wise description (see Section 2.3).
Following the parametrization and tying scheme of Setion 2.3, we do the spectral analysis for the individual pulses of the GRB. 
In the following analysis, we use the TTE data of NaI --- n0, n1, n2 and BGO --- b0 for our analysis. The 
constant, which takes care of the relative normalisation of the detectors should not vary throughout the burst. Hence, we freeze them to 
the values obtained in the time-integrated analysis, i.e., 2.25, 2.32, 2.34 and 3.24 respectively.
Additionally, we make the following changes compared to the time resolved spectral analysis 
discussed earlier. We divide the data into spectra of equal total counts rather than 
equal time bins so that equal importance is given to all individual spectra. Further, the
spectral data in each bin is regrouped into spectral channels to provide an uniform SNR.
We also note that the 30 to 40 keV region of the spectrum of this GRB has the known calibration issues
due to K-edge of NaI (see e.g., Guiriec et al. 2011). This does not matter much for parameter estimations, but,
if one wants to compare different models in terms of $\chi^2$ then it is wise to neglect these bins. In the following,
we have done the spectroscopy by neglecting the 30-40 keV band. 
\subsubsection{Analysis of Pulse 2}
This pulse constitutes the major portion of the burst. The count rate is $\gtrsim 3$ times higher than pulse 1. Hence, we can 
analyse this pulse with greater accuracy and later use our experience to analyse the other one. We perform the analysis for two 
cases as follows:

\textit{Case I --- Analysis for count per time bin $\gtrsim$ 3000}:
This analysis is done by dividing the second pulse from 17.0 s onwards, requiring $\gtrsim$ 3000 counts per time bin. We divide the 
pulse into two parts. 17 to 21.45 s is the rising part and the rest up to 40.45 s is the falling part.
In the rising part, we get 3 time bins and the falling part has 9 of them. The spectral bins in the energies \textgreater 100 keV sometimes 
show less than 2$\sigma$ count, while \textless 15 keV show less than 3$\sigma$ count. Hence, we merge the 8 keV to 15 keV bins to form 
1 bin; 100 to 900 keV bins are merged into 7 bins, with progressively higher binning at higher energies . Similarly, spectral bins of the BGO 
(200 keV to 30 MeV) are merged into 5 large bins. All the spectral fit parameters are listed in Table~\ref{p2}.

(a) \textit{Rising part}:
In the rising part (first four rows of Table~\ref{p2}), the BBPL, compared to the other models is inferior with 
$\chi^2$ (dof) --- 455.92 (354). If we parametrize only the BB norm, and treat the PL norm as free parameter, the corresponding
$\chi^2$ (dof) is 487.68 (354). A BBPL fitting with no constraint gives $\chi^2$ (dof) = 451.09 (348). While we have gained dof, the
$\chi^2_{red}$ remains of the same order, which confirms that the parametrization works.
In comparison to BBPL, mBBPL is a better fit with $\chi^2$ (dof) = 355.68 (353), and significance of 2.55 $\sigma$ (98.93\% confidence). Band 
is better than BBPL model, with $\chi^2$ (dof) = 364.41 (354), and significance of 2.37 $\sigma$ (98.23\% confidence). This suggests that the 
radiation mechanism in the rising part of the second pulse may be a photospheric emission, but the thermal part is a modified 
blackbody (mBB) rather than a simple BB. However, if we compare these values with a 2BBPL model, then we immediately see that 
this model is the best with $\chi^2$ (dof) = 351.78 (352). As the set of parameters of BBPL model is a subset of 2BBPL, the 
significance of 2BBPL compared to BBPL is much higher --- 9.29 $\sigma$ (100\% confidence). Compared to the Band model, 2BBPL model has a
significance of 0.86 $\sigma$ (60.94\% confidence), which shows that 2BBPL is only marginally better than Band model.

(b) \textit{Falling part}:
In the falling part (see Table~\ref{p2}), the Band model is better compared to the BBPL as well as the mBBPL model. Compared to
the mBBPL model, the Band model has 35.22 less $\chi^2$ with one more dof. A comparison with 2BBPL model, on the other hand, shows 
that 2BBPL is better than Band model at 1.03 $\sigma$ significance (69.71\% confidence). Compared to Band model, 2BBPL has 40.64
less $\chi^2$ with two less dof. The Band model does not show much difference in terms of residuals of individual spectral fit. 
But, it is only when we perform a parametrized joint fit, then we realise that the 2BBPL model is marginally better than Band model (Table 4).
Hence, in this region either Band or 2BBPL is the best model, with 2BBPL marginally better than Band model.
In Figure~\ref{spectrum}, we have shown the significance of 2BBPL fitting over the BBPL fitting as a case study. The residual of
BBPL model shows excess at various channels. No such structure is visible in the residual of 2BBPL fit. Note that the NaI K-edge is present
in both the residuals between 30-40 keV. We have done the fitting both by including and excluding this band. When all the channels are used
the $\chi^2$ (dof) of BBPL and 2BBPL are 340.85 (217) and 239.82 (215) respectively. 2BBPL is preferred over BBPL at a significance of
8.42 $\sigma$ (100\% confidence, p=$3.86\times10^{-17}$). If we exclude the 30-40 keV bins, the corresponding $\chi^2$ (dof) are
300.26 (197) and 193.17 (195), while 2BBPL is preferred at a significance of 
9.01 $\sigma$ (100\% confidence, p=$2.10\times10^{-19}$)

\textit{Case II --- Analysis for count per time bin $\gtrsim$ 1000}:
To check whether lowering the size of the time bins improves the BBPL fitting, we perform the same analysis for count per time bin
$\gtrsim$ 1000. As the count rate is lower, we merge the 100 to 900 keV of NaI detectors into 5 channels rather than 7. The rest of the
binning remains the same. We obtain 10 bins in the rising and 29 bins in the falling part.

(a) \textit{Rising part}: 
The values of $\chi^2$ (dof) for BBPL, Band, mBBPL and 2BBPL are 1247.91 (1187), 1328.50 (1187), 1239.20 (1186) and 1222.76 (1185), 
respectively. Compared to BBPL model, the Band model is preferred at 1.36 $\sigma$ (82.76\% confidence), mBBPL is preferred
at 1.56 $\sigma$ (88.17\% confidence), while 2BBPL is preferred at 9.66 $\sigma$ (100\% confidence). 2BBPL is preferred over Band
model at 0.89 $\sigma$ (62.60\% confidence). Hence, the conclusions of Case I remains unaltered. 

(b) \textit{Falling part}:
Similarly, in the falling part, the finer bin makes equal impact on all the models and hence, the conclusion remains the same.
Note that compared to mBBPL, 2BBPL model has 113.84 less $\chi^2$, with the cost of one more dof. Hence, 2BBPL model is preferred 
over mBBPL in the falling part (1.31 $\sigma$ with 68.99\% confidence). Compared to BBPL model, Band, mBBPL and 2BBPL are 
preferred at 3.12 $\sigma$ (99.82\% confidence), 2.67 $\sigma$ (99.24\% confidence) and 19.61 $\sigma$ (100\% confidence), respectively.
2BBPL is marginally better than Band at 0.95 $\sigma$ (65.64\%).

\subsubsection{Analysis of Pulse 1}
The time-resolved spectra of this pulse are extracted by requiring $\gtrsim1000$ counts per bin, as the photon count is $\sim 1/3$rd of 
pulse 2. As before the spectral bins of NaI are binned in 8-15 keV and 100-900 keV, while BGO spectral channels are merged to form 5 broad
channels. The 30-40 keV band is neglected. The results of different fits are reported in Table~\ref{p1}. 

(a) \textit{Rising part}: The rising part has only one bin from -1.0 to 2.15 s. Interestingly the BBPL model is marginally better
than the Band model in the rising part (see Table~\ref{p1}) at 0.87 $\sigma$ (61.46\% confidence). The mBBPL model has comparable $\chi^2$ as 
BBPL, but with one more parameter. 2BBPl model has a significance of 1.04 $\sigma$ (70.17\% confidence) compared to Band model, while
the same model has a significance of 2.19 $\sigma$ (97.12\% confidence) compared to BBPL model.

(b) \textit{Falling part}: In the falling part of this pulse, the BBPL model is no longer the best model. The Band is the best model
with $\chi^2$ (dof) = 544.65 (473). mBBPL and 2BBPL models are comparable to Band with $\chi^2$ (dof) 548.22 (472) and 544.79 (471).
Hence, the spectrum may be still thermal, though the thermal part is no longer a simple BB, but either a multi-color BB (mBB) or
has multiple spectral component (one more BB) or simply synchrotron dominated (Band). Note also that the low energy
photon index ($\alpha=-0.86_{-0.19}^{+0.22}$), in the falling part is within the regime of synchrotron model, which is clearly
in contrast with the rising part ($\alpha=-0.55_{-0.22}^{+0.26}$). This phenomenon of softening of photon index at the falling part
of a pulse can be seen for all the pulses (see Table~\ref{p2} and ~\ref{p1}). 
In Table~\ref{sigma_level}, we have listed all the significance levels (in terms of p-value, sigma level and confidence level --- C.L.) 
of a model over another. Model$_1$ is the primary model, while
Model$_2$ is the alternative model. It is clear from the table that 2BBPL model is preferred over Band model, though marginally, in some
cases. The p values denote the probably that the alternative hypothesis is incorrect. Hence, lower the value of p, better is the alternative
model over the primary model. Only in one case, namely the falling part of pulse 1, Band is preferred over 2BBPL. But the p-value of this case
is 0.48, which signifies that they are only comparable. Interestingly, if we use finer bin size, the significance of Band and mBBPL 
over BBPL model decreases in the second pulse. The significance of 2BBPL, however, increases.

\subsubsection{Connection between the rising and the falling part}
The smooth variations of the parameters demand that the temperature, kT (of BBPL or mBBPL or 2BBPL) or peak energy, $E_{peak}$ (of Band model)
should be continuous functions of time, even during the pulse peak time. Hence, these parameters should match at the peak 
within errors to the one predicted by the empirical law. We follow the evolution law of the rising part to predict these 
parameters in the first bin of the falling part. We compare these values with the corresponding observed values. In Figure~\ref{connection},
we have plotted the observed values with respect to the predicted values. Note that the error bars of the observed values are much 
less compared to those of the predicted values. The sources of errors in the predicted values are errors in the evolution parameter, 
$\mu$ and the errors in the actual parameter at the starting bin of the rising part. Generally, the parameter $\mu$ has large errors, 
which affects the errors of the predicted values considerably. The data points are essentially the same for both wider bin (open circles) 
and finer bin (filled circles). The dot-dashed line, which shows the equality of the observation and prediction, goes through all the points.

\subsection{Comparison with GRB 090618}
GRB 090618 is an interesting GRB in many aspects (for details see Ghirlanda et al. 2010; Rao et al. 2011; Basak \& Rao
2012a). This is a GRB with very high fluence. The redshift (z) of this object is 0.54
and the 
total fluence is 
$3398.1 \pm 62.0 \times 10^{-7}$ erg cm$^{-2}$, when integrated over its duration (182.27 s).
In terms of fluence, this is the brightest among all GRBs detected by Fermi. This is a very long GRB with multiple peaks. 
It has four broad pulses as follows. Pulse 1 is rather a clean precursor from -1 s to 40 s. The second pulse is well separated 
from this precursor, and occurs from 50 s to 75 s, with two structures in 50 to 61 s. The third pulse, occurring from 75 s to 100 s, 
is contaminated with 
the falling part of the second pulse, and the rising part of the fourth pulse. The fourth pulse occurs from 100 s to 124 s. Though 
the secondary pulses (i.e., other than the precursor) are sometimes overlapping, we can still examine the spectral variation
in the first pulse and in the major portions of the other pulses. GRB 081221 has only one secondary pulse, which makes it
more convenient. However, in contrast with GRB 090618, the precursor of GRB 081221 has overlaps with the secondary pulse.
In order to compare the results of GRB 081221, we shall take the precursor and the second pulse of GRB 090618.

\subsubsection{Precursor pulse}
The time-resolved spectra of this pulse are obtained by requiring minimum of 1000 counts per bin. We obtain 10 spectra in the 
rising part (-1.0 to 14.15 s) and 11 spectra in the falling part (14.15 to 40.85 s). For the rising part, the parameters, $\mu$
and $\nu$ are obtained by assuming the start time at -10.0 s (see Section 2.3). The results of spectral fitting by different
models are reported in Table~\ref{090618_a}. The advantage of this pulse 1 over 081221 is it is longer and brighter, enabling
us to parametrize the rising part. Also, this pulse is fully separated from the secondary events. It is clear from Table~\ref{090618_a}
that BBPL model is inferior to the  mBBPL model for this pulse. In case of GRB 081221, we found that the BBPL and mBBPL models are 
comparable to each other, and marginally better than Band model. In this case, we definitely need a mBBPL rather than a BBPL for comparable
fit as Band. Note that 2BBPL model is only comparable, but not better than mBBPL. Hence, the spectrum in the rising part may be
modified blackbody dominated. In the falling part, mBBPL is again comparable to Band model. The 2BBPL model is superior
than all the models in the falling part. The same conclusion was drawn for GRB 081221. Hence, there is hardly any difference of 
spectral evolution in the precursor pulse between these two GRBs.

\subsubsection{The second pulse}
This pulse is more difficult to analyse, because, it has two small structures in the rising part and two pulses at the peak.
Hence, we ignore the 50 to 61 s of data and analyse only 61 to 75 s. The falling part covers 64.35 to 74.95 s, where we obtain 24 
time-resolved spectra requiring 2000 counts per bin. This, in principle, can be lowered as we have higher count rate, but, following
the analysis of GRB 081221, we restrict ourselves to a moderate count per bin. The results of joint spectral fit with
various models are reported in Table~\ref{090618_b}. It is clear from this table that BBPL and mBBPL are not the correct models
in the falling part. In comparison Band model is much better. However, the 2BBPL model, which shows comparable or better fit than Band
in all cases, is only comparable to mBBPL in this particular case. This may arise due to the fact that this pulse is actually a 
combination of two highly overlapping pulses (see Rao et al. 2011). Hence, we redo the analysis on 11 spectra from 69.25 s to 
74.95 s, which covers only the falling part of the second pulse. We obtain the following $\chi^2$ (dof): 1137.67 (991),
1179.13 (990) and 1156.25 (989) for Band, mBBPL and 2BBPL, respectively. Hence, the Band and 2BBPL models are comparable,
though we cannot rule out the possibility of contamination even in this falling part.

In the rising part (see Table~\ref{090618_b}), though mBBPL is better than BBPL, it is inferior to   2BBPL and Band. Hence, 
the rising part of this pulse is probably synchrotron dominated. 2BBPL is comparable to Band. In summary of this pulse,
the whole episode can be described by Band model. This is in contrast with the second pulse of GRB 081221, where mBBPL
clearly dominates the rising part, and then it is taken over by Band. 

\section{Discussion and Conclusions}
From purely data analysis point of view, there are essentially two parts in a GRB spectrum --- the peak of the spectrum and 
the wings, which extend to very low and very high energies. The fundamental difference between a Band only model and the 
models with thermal and non-thermal parts is: while Band accounts for the peak position, with exponential fall in the wings, 
the other models have thermal component to account for the peak position, and a powerlaw, falling slower than an exponent,
holding the spectra at the wings. In principle, all these spectra should be equally good at the peak position, except 
for the fact that Band and mBB have broader peak than simple BB, while 2BBPL has double hump. Hence, the difference between 
these models arises mainly in the wings. The BB is inferior to the others if the peak is not narrow. As photon count at the peak 
is larger, the residual should show up immediately. But, this is not easy to see if the difference occurs in the wings. Consequently, 
the three very different models, namely, Band, mBBPL and 2BBPL show comparable $\chi^2$, while fitting time-resolved data. Hence, 
re-binning at these wings plays a very important role to pin down the correct model. However, we cannot expect an order of magnitude 
improvement in the $\chi^2$, because, binning in these wings gives 6-8 broad bins (see Section 3.3.1) with large errors, while 
the major statistics comes from the peak position.

 We have found, in our analysis, that the spectrum changes from one pulse to
the other, and even within a pulse. The fact that one of these four models is superior than the others, in a particular episode,
points to the fundamental radiation mechanism. We see that this change of superiority is not random. For example, the first pulse
of both the GRBs have shown that a mBBPL model is better, though marginally, than Band in the rising part. For GRB 081221, this 
could be described even by BBPL model in the rising part, which is really pointing towards the thermal origin of radiation in 
the first pulses. Similar observations are reported in the literature e.g., Ryde et al. (2009) showed, in the 1-3 s 
time bin of BATSE detected GRB 981021, that a BBPL is better fit than a Band model. Note that the low energy photon index ($\alpha$) 
of the Band model crosses the synchrotron limit in the rising part, where the thermal models are adequate. However, in the falling
part of all the pulses, where Band is better than mBBPL, $\alpha$ is consistent with the synchrotron limit. Hence, we can safely
conclude that the radiation mechanism starts with a thermal origin, but is rapidly overtaken by synchrotron
mechanism. The first pulse may be dominated by the photospheric emission in all episode, but the second pulse is mostly synchrotron dominated.
The second pulse may or may not have a thermal origin. For example, the second pulse of GRB 081221 shows a mBBPL model in the rising part,
which then becomes synchrotron dominated in the falling part. On the other hand, the second pulse of 090618 is always synchrotron 
dominated. Hence, the transition between these different radiation paradigms is smooth and repeatable. 

\begin{figure}\centering
{

\includegraphics[width=6.5in]{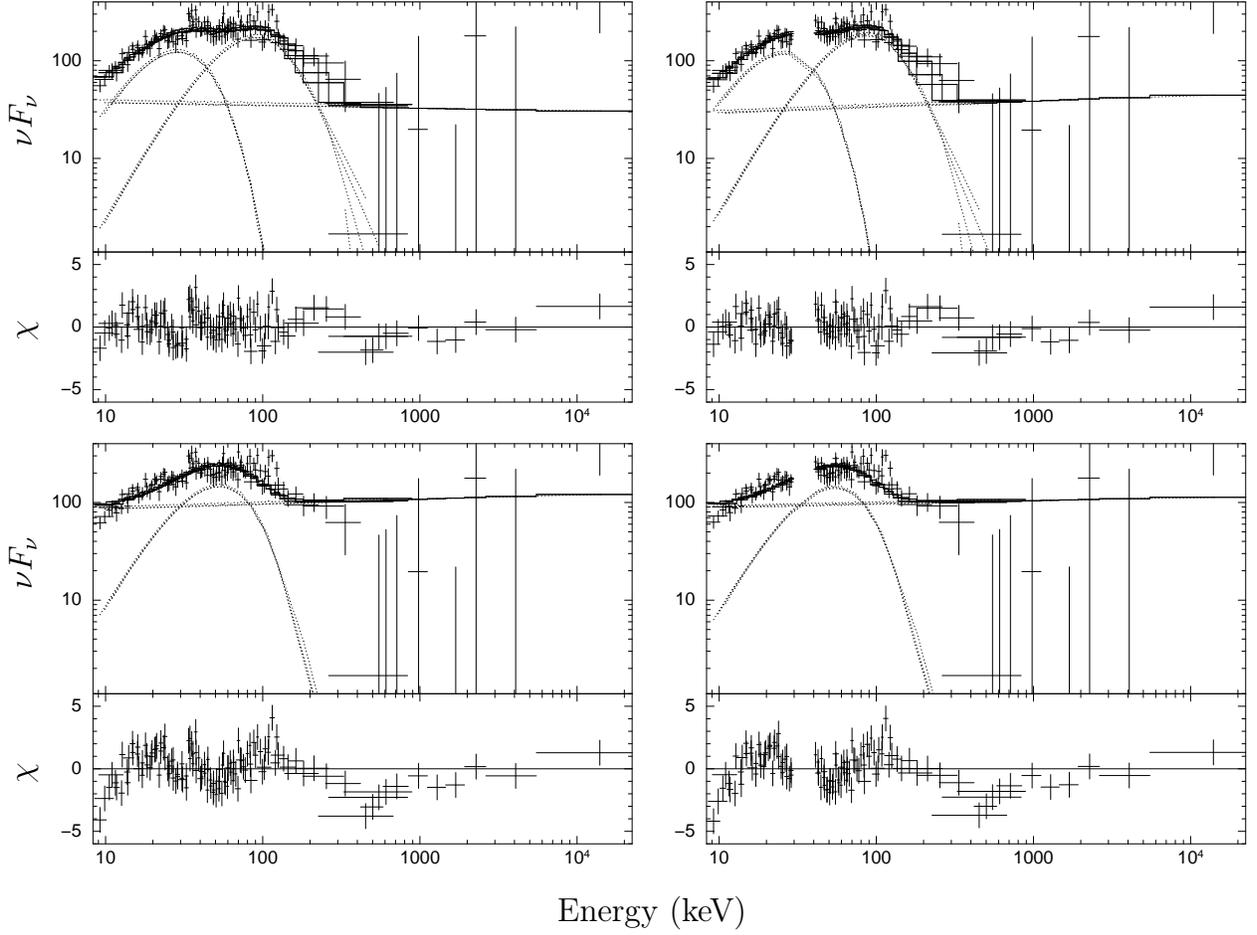} 

}
\caption{Fitting the 29.0 to 32.0 s spectral data of GRB~081221 with BBPL (lower panels) and 2BBPL (upper panels) models with the 30-40 keV band included (left panels) and 
neglected (right panels). $\nu F_{\nu}$ has the unit keV$^2$(Photon cm$^{-2}$ s$^{-1}$).
Note the double hump structure in the residuals of the BBPL
model. This structure is taken care of by the 2BBPL model.
 }
\label{spectrum}
\end{figure}

\begin{figure}\centering
{

\includegraphics[width=6.5in]{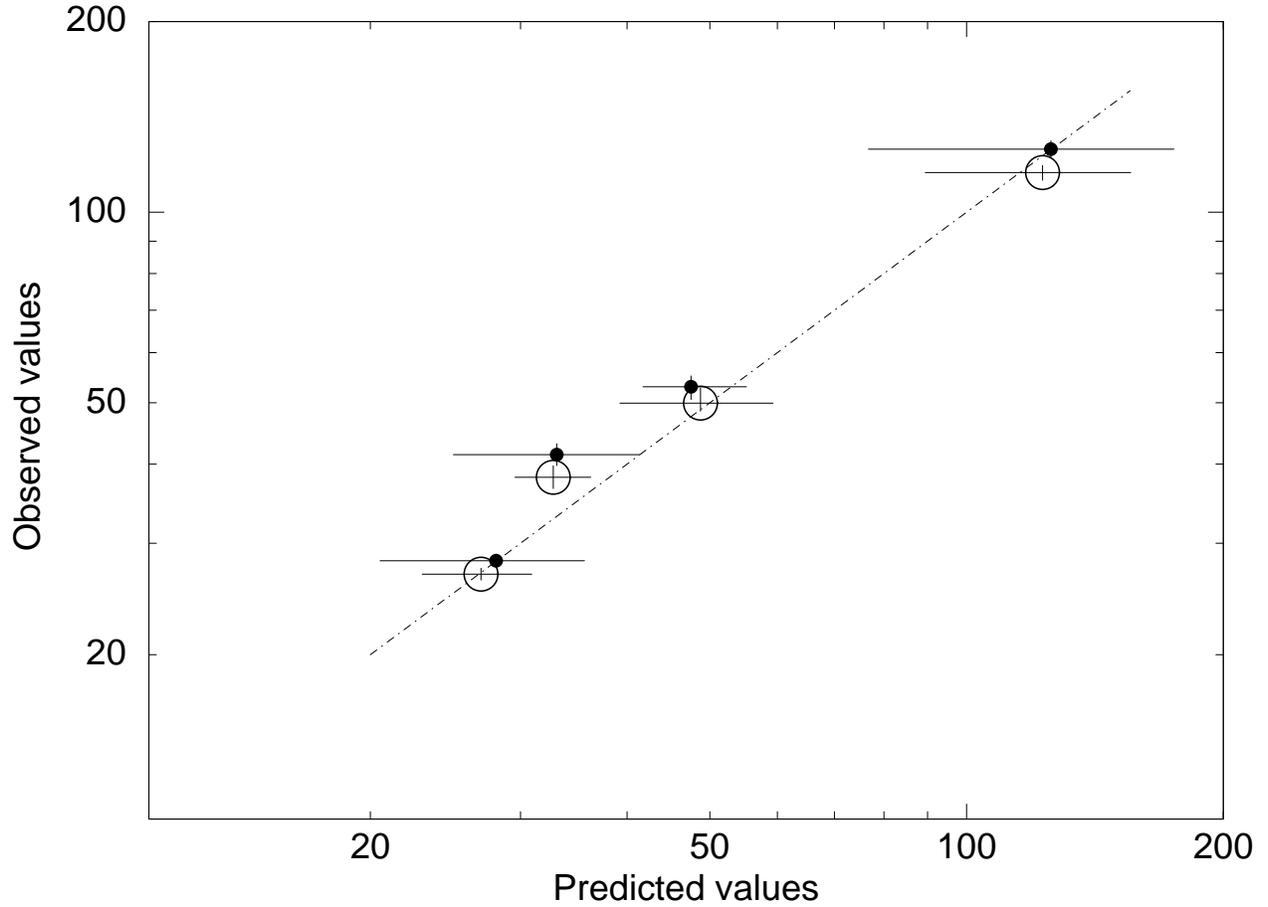} 

}
\caption{Predicted peak energy ($E_{peak}$) of Band, $kT$ of BBPL, $kT_{in}$ of mBBPL and $kT_h$ of 2BBPL models are
compared with the observed values. The open circles are the values obtained for wider bin size ($\gtrsim 3000$ counts per bin), 
while the filled circles represent values obtained for finer bin size ($\gtrsim 1000$ counts per bin). The dot-dashed line is the
line of equality.
}
\label{connection}
\end{figure}

\begin{figure}\centering
{

\includegraphics[width=6.5in]{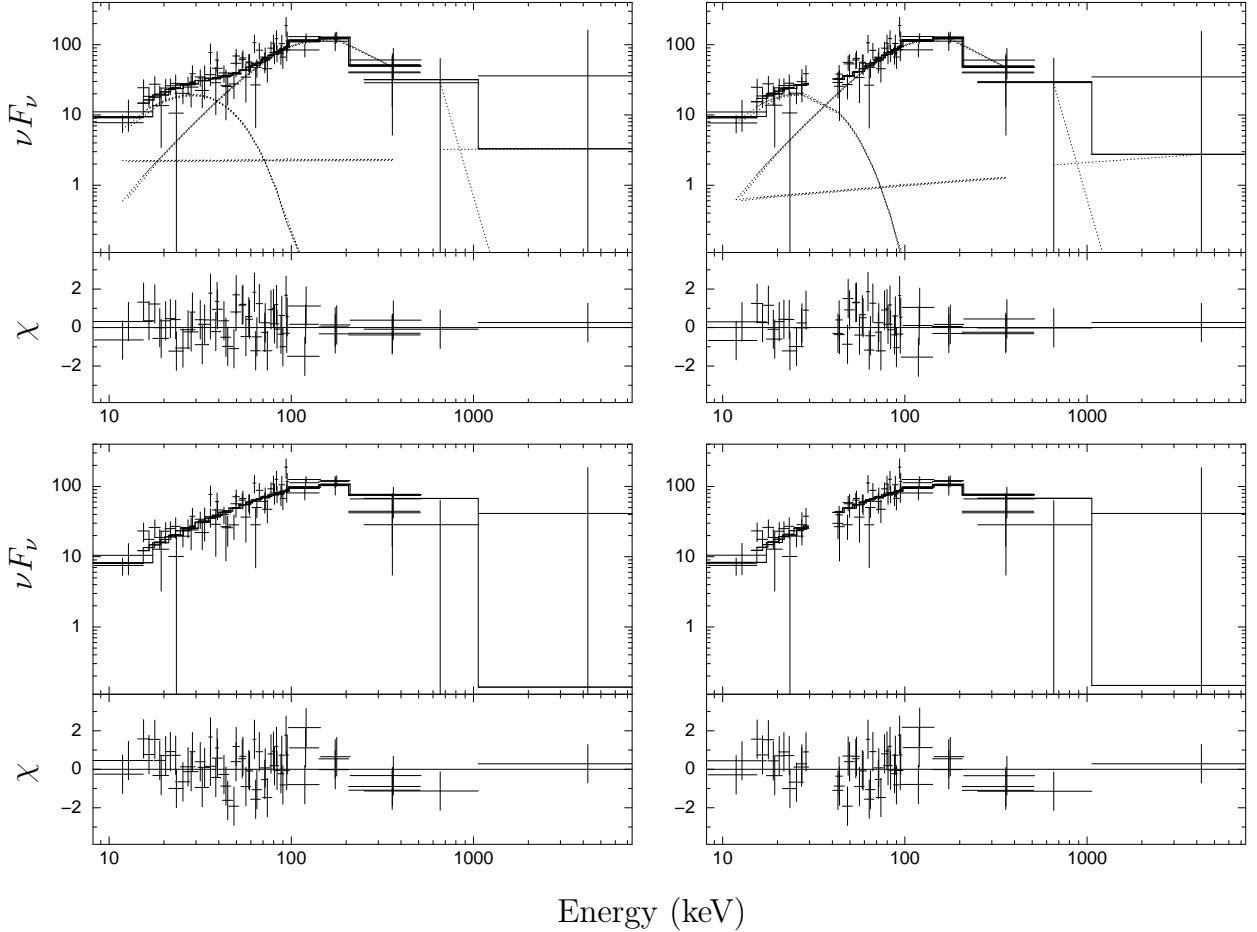} 

}
\caption{Comparison of spectral fitting between 2BBPL (upper panels) and Band (lower panels) model for -1.0 to 2.15 s time bin of GRB~081221. 
$\nu F_{\nu}$ has the unit keV$^2$(Photon cm$^{-2}$ s$^{-1}$).
The right panels show the fit with the 30-40 keV channels neglected. Note the structure in the residual of the Band model ---
positive excess near 15 keV and 150 keV, and negative excess near 40-60 keV. Compared to this the 2BBPL model does not show any structures 
in the residual. The 2BBPL model is preferred over the Band model at 1.04 $\sigma$ with 70\% confidence level and p-value = 0.298 for 30-40 keV neglected
case and at 0.95 $\sigma$ with 65.5\% confidence level and p-value = 0.341 for 30-40 keV included case, based on F-tests.
 }
\label{spectrum_compare}
\end{figure}

In comparison to the mBBPL model, we note that the 2BBPL is particularly better in all episodes. This model
sometimes shows superiority to the  Band model, even at the falling part of a pulse, except for the second pulse of GRB 090618, 
though we cannot rule out the possibility of two highly overlapping pulses in this particular case. Softer component than Band were 
reported  for a few BATSE GRBs by Preece et al. (1996). Shirasaki et al. (2008), using the time-resolved spectral data of 
GRB 041006, detected by HETE-2 (2 keV to 400 keV), found multiple spectral components, each having characteristic evolution. 
After the launch of Fermi satellite, these earlier claims were reconfirmed in some cases. For example, Guiriec et al. (2011), fitting the 
time-integrated spectrum of GRB 100724B, have shown the presence of an additional blackbody (BB) component along with the traditional 
Band spectrum (also see Burgess et al. 2011). In our analysis, we have used two blackbodies to account for the softer components.
The origin of these two components is speculative. They might be different locations of the boosted front of the fireball having same
temperature, but different boosting factors. Alternatively, they can be different seed photon baths, up-scattered by the bulk material.
Irrespective of its origin, this model shows superiority to all other models in all episodes. Note that, though 2BBPL model has double
hump in the peak, one of the peaks may occur in the lower wing (i.e., \textless 15 keV). Hence, it is easy to identify this model, only if
the difference occurs at the peak.
Figure~\ref{spectrum} clearly shows the double hump in the residuals of the BBPL fit. Hence, it is easy to visualise the 2BBPL model from 
this figure. The  Band model, however, has similar residuals as the 2BBPL model. Hence,
the  data are not sufficient to distinguish between these two models,
except when we perform a parametrized joint fit. 
In Figure~\ref{spectrum_compare}, we have shown the marginal superiority of the 2BBPL fit over the Band model as a case study
of the rising part of the first pulse (-1.0 to 2.15 s) of GRB 081221. In the right panels, we have plotted the fitted data neglecting the 30-40 keV
channels. The upper panels are 2BBPL fits, while the lower panels are Band model fits. Residuals of the Band model shows structures
with excesses in 15 keV, 50-60 keV, and 150 keV regions. These are not present in the residuals of the 2BBPL model. Of course, the difference is not as
prominent as the case of Figure~\ref{spectrum}. The 2BBPL model is preferred over the Band
model at 1.04 $\sigma$ (p=0.298, 70.17\% confidence). 
Hence, 2BBPL is only marginally better than the Band model.

\begin{figure}\centering
{

\includegraphics[width=5.0in]{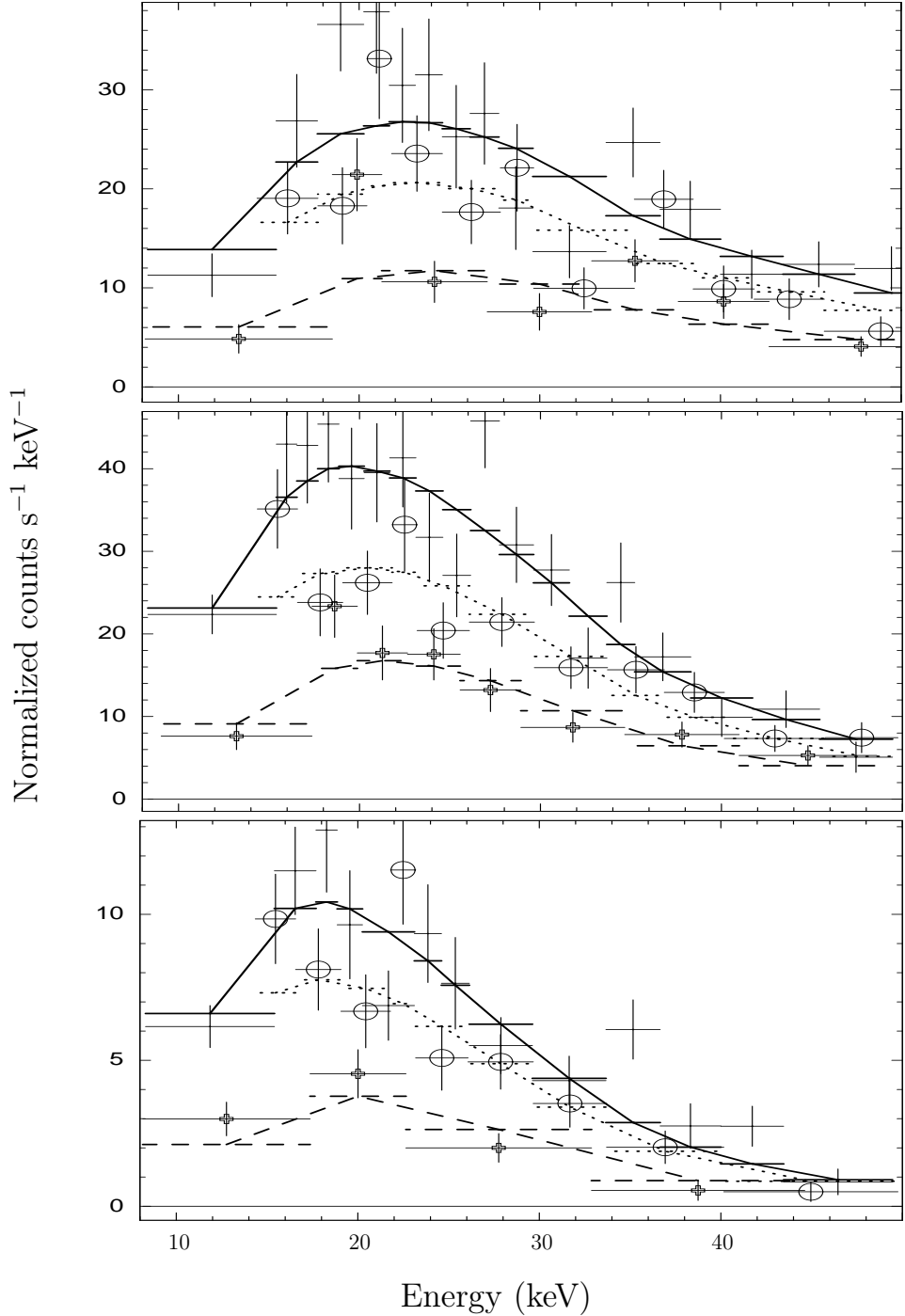} 

}
\caption{The residual in various time bins of the falling part of GRB 081221. The data used are those with $\gtrsim 3000$ counts per bin.
The time bins used are 2nd, 6th and 9th bins (from top to bottom). The residuals are obtained by omiting the lower BB from the 2BBPL fit.
The lower BB models are overplotted with the residuals to show the significance of this BB component. Different lines and symbols show 
different detectors. Note that the lower BB temperature shifts to the lower energy with time.
 }
\label{lower_bb}
\end{figure}

To visualise the evolution of the lower black body component, we have plotted in Figure~\ref{lower_bb}, the residuals
of 2BBPL fit, with the lower BB omitted. This technique is well known for finding iron line profile in the inner accretion disks
of black holes (see e.g., Miller 2007). We fit the spectrum with the 2BBPL model and then omit the lower BB. The residual (expressed as 
normalized counts keV$^{-1}$ s$^{-1}$) of the fit clearly shows this BB component. Residuals of different detectors are shown by different 
markers. We have overplotted the lower BB model (in terms of normalized  counts keV$^{-1}$ s$^{-1}$) on the residual to guide the eye.
We have plotted these residuals for second, sixth and ninth time bins from top to bottom panels to show that the BB peaks at lower 
energies at later times.

In summary, we have rigorously used the evolution of parameters in the pulses of a GRB to construct various spectral
models with a minimum number of parameters. We have constructed Band model with parametrized peak evolution and tied photon indices,
BBPL with parametrized norm ratio of the BB and PL, parametrized temperature, and tied PL indices. Apart from these, we have used
mBBPL and 2BBPL, which, other than the parametrizations of BBPL, have tied p indices, and tied ratios of temperatures and norms, respectively.
This new method is quite general in the sense that any such model can be incorporated with suitable parametrization.
The fact that the parametrization works demands a close look into the theoretical predictions of various radiation models.
These models, irrespective of their complexities, should produce such smooth variations of parameters within a pulse of a GRB.
Also, if there is really a transition from one radiation mechanism to another, one should correctly model the mechanism
of such transition. The fact that the synchrotron model is applicable at the falling part of the pulses,
without invoking any other component, is really intriguing and demands a close look at the predictions of the internal shock model.

One of the surprising result obtained in this work is that the 2BBPL model is statistically superior to the
other models in most of the episodes in these two GRBs. Basak \& Rao (2012c) have used this model for GRB 090902B. 
The residual of BBPL fit clearly shows double humps (see Figure 3 of Basak \& Rao 2012c) which are taken care by the 
two peaks of the 2BBPL model. We selected this model purely in a phenomenological and data analysis perspective: 
distinct blackbody components (apart from the main peak in the spectrum) are seen in a few GRBs and while looking 
at the residuals, two humps are clearly discernible in a few time bins. 
Since these two are the brightest GRBs for such analysis (GRB 081221 is the brightest GRB
in the category of GRBs with single/ separable pulses and GRB 090618 is the brightest GRB in the Fermi era),
it is unlikely that we can reinforce this result by analysing data from other GRBs. One method could be to
get the pulse-wise spectral parameters of a sample of GRBs and relate them to other properties
of GRBs like redshift, afterglow properties etc. This will not only help us to identify the most 
appropriate spectral description but also to identify the emission mechanism operating during the
prompt emission.

\section*{Acknowledgments} This research has made use of data obtained through the
HEASARC Online Service, provided by the NASA/GSFC, in support of NASA High Energy
Astrophysics Programs.

\clearpage
\begin{longtable}{c|c|c|c|c}

\caption{Results of time-integrated spectral analysis of the GRBs.} \\

\hline
\endhead
\hline \multicolumn{4}{r}{\textit{Continued on next page}} \\
\endfoot
\endlastfoot

\hline
  GRB   & $t_1$, $t_2$ & \multicolumn{2}{c|}{This work} & Nava et al. (2011) \\
\cline{3-4}
(Model) &                & C-stat & $\chi^2$         &                    \\
\hline
\hline

 080904 & -4.096, 21.504    & $\alpha$=-1.22$^{+0.21}_{-0.20}$      & $\alpha$=-1.21$^{+0.20}_{-0.19}$   &  $\alpha$=-1.14$\pm0.05^a$  \\
 (CPL)  &                   & $E_p=40.1^{+3.92}_{-3.56}$            & $E_p=39.8^{+3.68}_{-3.34}$         &  $E_p=39.24\pm 0.75$ \\
        &                   & $\rm C^b$=1.08 (597)                  & $\chi^2_{red}$=1.23 (597)          &  C=1.14(587) \\

\hline 

 080925 & -3.840, 32.0      & $\alpha$=-1.06$^{+0.11}_{-0.10}$       & $\alpha$=-1.06$^{+0.11}_{-0.10}$   &  $\alpha$=-1.03$\pm0.03$   \\
 (Band) &                   & $\beta=$=-2.34$^{+0.30}_{-1.13}$       & $\beta=$=-2.24$^{+0.24}_{-0.74}$   &  $\beta=$=-2.29$\pm0.08$   \\
        &                   & $E_p=158.9^{+31.6}_{-24.4}$            & $E_p=157.3^{+33.5}_{-24.9}$        &  $E_p=156.8\pm 7.07$ \\
        &                   & C=1.17 (712)                           & $\chi^2_{red}$=1.14 (712)          &  C=1.13(716) \\

\hline 

 081118 & 0.003, 19.968     & $\alpha$=-0.42$^{+0.70}_{-0.48}$       & $\alpha$=-0.37$^{+0.70}_{-0.49}$   &  $\alpha$=-0.46$\pm0.10$   \\
 (Band) &                   & $\beta=$=-2.18$^{+0.16}_{-0.35}$       & $\beta=$=-2.14$^{+0.15}_{-0.19}$   &  $\beta=$=-2.29$\pm0.05$   \\
        &                   & $E_p=55.93^{+22.2}_{-12.5}$            & $E_p=54.0^{+19.7}_{-12.0}$         &  $E_p=56.79\pm 2.77$ \\
        &                   & C=1.17 (716)                           & $\chi^2_{red}$=1.02 (716)          &  C=1.16(601) \\

\hline 

 081207 & 0.003, 103.426    & $\alpha$=-0.58$^{+0.10}_{-0.09}$       & $\alpha$=-0.58$^{+0.12}_{-0.11}$   &  $\alpha$=-0.58$\pm0.02$   \\
 (Band) &                   & $\beta=$=-2.15$^{+0.17}_{-0.33}$       & $\beta=$=-2.13$^{+0.20}_{-0.41}$   &  $\beta=$=-2.22$\pm0.7$   \\
        &                   & $E_p=363.4^{+70.7}_{-51.5}$            & $E_p=364.5^{+82.8}_{-59.4}$        &  $E_p=375.1\pm 13.2$ \\
        &                   & C=1.43 (713)                           & $\chi^2_{red}$=1.02 (713)          &  C=1.74(596) \\

\hline

 081217 & -28.672, 29.696   & $\alpha$=-1.09$^{+   0.15}_{  -0.14}$  & $\alpha$=-1.10$^{+0.16}_{-0.14}$   &  $\alpha$=-1.05$\pm0.04$  \\
 (CPL)  &                   & $E_p=193.0^{+65.9}_{-37.3}$            & $E_p=200.5^{+77.2}_{-41.7}$        &  $E_p=189.7\pm 11.2$ \\
        &                   & C=1.19 (715)                           & $\chi^2_{red}$=1.06 (715)          &  C=1.46(599) \\

\hline 

 081221 & 0.003, 39.425     & $\alpha$=-0.84$^{+0.06}_{-0.05}$       & $\alpha$=-0.84$^{+0.06}_{-0.06}$   &  $\alpha$=-0.82$\pm0.01$   \\
 (Band) &                   & $\beta=$=-4.24$^{+0.93}_{-10.2}$       & $\beta=$=-3.89$^{+0.69}_{-7.1}$    &  $\beta=$=-3.73$\pm0.20$   \\
        &                   & $E_p=85.25^{+2.89}_{-3.08}$            & $E_p=85.09^{+3.23}_{-3.19}$         &  $E_p=85.86\pm 0.74$ \\

        &                   & C=1.64 (595)                           & $\chi^2_{red}$=1.49 (595)          &  C=1.67(600) \\

\hline 

 081222 & -0.768, 20.736    & $\alpha$=-0.89$^{+0.14}_{-0.12}$       & $\alpha$=-0.89$^{+0.14}_{-0.12}$   &  $\alpha$=-0.90$\pm0.03$   \\
 (Band) &                   & $\beta=$=-2.46$^{+0.37}_{-1.37}$       & $\beta=$=-2.32$^{+0.31}_{-0.98}$   &  $\beta=$=-2.33$\pm0.10$   \\
        &                   & $E_p=169.2^{+37.3}_{-27.4}$            & $E_p=168.9^{+39.1}_{-29.8}$        &  $E_p=167.2\pm 8.28$ \\
        &                   & C=1.12 (595)                           & $\chi^2_{red}$=1.07 (595)          &  C=1.23(604) \\

\hline

 090129 & -0.256, 16.128    & $\alpha$=-1.43$^{+   0.19}_{  -0.16}$  & $\alpha$=-1.46$^{+0.18}_{-0.16}$   &  $\alpha$=-1.46$\pm0.04$  \\
 (CPL)  &                   & $E_p=170.4^{+130.0}_{-48.5}$           & $E_p=195.5^{+212}_{-63.5}$         &  $E_p=166.0\pm 15.1$ \\
        &                   & C=1.09 (596)                           & $\chi^2_{red}$=1.03 (596)          &  C=1.12(602) \\

\hline

 090709 & 0.003, 18.432     & $\alpha$=-1.04$^{+   0.38}_{  -0.32}$  & $\alpha$=-1.08$^{+0.37}_{-0.31}$   &  $\alpha$=-0.96$\pm0.08$  \\
 (CPL)  &                   & $E_p=116.7^{+76.9}_{-30.6}$            & $E_p=124.1^{+101}_{-34.7}$         &  $E_p=137.5\pm 12.5$ \\
        &                   & C=1.05 (596)                           & $\chi^2_{red}$=1.01 (596)          &  C=1.17(602) \\

\hline

 091020 & -3.584, 25.088    & $\alpha$=-1.31$^{+   0.29}_{  -0.18}$  & $\alpha$=-1.32$^{+0.22}_{-0.19}$   &  $\alpha$=-1.20$\pm0.06$  \\
 (CPL$\rm )^c$  &                   & $E_p=255.7^{+332.0}_{-92.0}$           & $E_p=276.4^{+485.0}_{-107.0}$      &  $\beta=$=-2.29$\pm0.18$  \\
        &                   & C=1.03 (354)                           & $\chi^2_{red}$=0.95 (354)          &  $E_p=186.8\pm 24.8$  \\
        &                   &                                        &                                    &  C=1.18(354) \\

\hline 

 091221 & -2.048, 37.889    & $\alpha$=-0.62$^{+0.27}_{-0.21}$       & $\alpha$=-0.62$^{+0.34}_{-0.23}$   &  $\alpha$=-0.57$\pm0.05$   \\
 (Band) &                   & $\beta=$=-2.40$^{+0.50}_{-3.15}$       & $\beta=$=-2.26$^{+0.45}_{-2.80}$   &  $\beta=$=-2.22$\pm0.10$   \\
        &                   & $E_p=191.3^{+67.4}_{-47.5}$            & $E_p=189.5^{+76.8}_{-57.1}$        &  $E_p=194.9\pm 11.6$ \\
        &                   & C=1.42 (474)                           & $\chi^2_{red}$=1.12 (474)          &  C=1.44(466) \\

\hline



\end{longtable}

\begin{tiny} $^a$ The errors quoted from Nava et al. (2011) are symmetric errors. Errors for this work are 3$\sigma$ errors. 
$^b$ C is the reduced C-stat value, the number in the parentheses are dof.
$^c$ The Band spectrum showed unbound 3$\sigma$ errors, we found better fit with CPL for this GRB
\end{tiny}
\clearpage

\thispagestyle{empty}

\begin{sidewaystable}
\begin{scriptsize}
\caption{Results of BBPL and Band fitting of time-resolved data of GRB 081221. For BBPL, values are quoted for both
powerlaw index ($\Gamma$) free and frozen to mean value at the high count rate region --- 1.83.
K$_1$ is BB normalization, while K$_2$ is that of the PL. The norms should be used 
as the relative normalization, as there is another constant multiplication due to detector effective area.
Bin \#s are 0-13, with 0 denoting -1 to 2 s and subsequently equal bin size of 3 s is applied.
Errors in K$_2$ of the bin \# 13 of BBPL ($\Gamma$ free) could not be determined. In many cases,
only the upper error in $\beta$ could be determined.}

\hspace{-6.5cm}

\begin{tabular}{c|ccccc|cccc|cccc}

\hline

Bin & \multicolumn{5}{c|}{BBPL ($\Gamma$ free)} & \multicolumn{4}{c|}{BBPL ($\Gamma$ frozen to 1.83)} & \multicolumn{4}{c}{Band} \\
\cline{2-14}
\# & kT & K$_1$ & $\Gamma$ & K$_2$ & $\chi^{2}_{red}(dof)$ & kT & K$_1$ & K$_2$ & $\chi^{2}_{red}(dof)$ & $\alpha$ & $\beta$ & $E_{peak}$ & $\chi^{2}_{red}(dof)$\\
\hline
\hline
0 & $38.03_{-4.28}^{+4.91}$ & $2.99_{-0.68}^{+0.67}$ & $1.73_{-0.25}^{+0.42}$ & $5.31_{-3.15}^{+12.72}$ & $1.03(67)$ & $38.28_{-3.97}^{+4.78}$ & $3.14_{-0.46}^{+0.49}$ & $7.18_{-2.16}^{+2.25}$ & $1.01(68)$ & $-0.28_{-0.30}^{+0.36}$ & $-10.0$ & $178.06_{-27.57}^{+42.39}$ & $1.03(67)$ \\
1 & $16.26_{-2.21}^{+2.45}$ & $1.76_{-0.36}^{+0.39}$ & $1.77_{-0.13}^{+0.15}$ & $14.27_{-5.98}^{+9.51}$ & $1.34(76)$ & $16.92_{-1.63}^{+1.87}$ & $1.83_{-0.33}^{+0.34}$ & $17.33_{-2.99}^{+3.08}$ & $1.33(77)$ & $-0.69_{-0.22}^{+0.39}$ & $-3.76_{-\infty}^{+1.30}$ & $77.92_{-15.58}^{+11.08}$ & $1.10(76)$ \\
2 & $10.14_{-2.07}^{+3.12}$ & $0.78_{-0.27}^{+0.28}$ & $1.99_{-0.23}^{+0.27}$ & $20.66_{-13.36}^{+28.87}$ & $1.02(67)$ & $9.07_{-1.33}^{+1.49}$ & $0.84_{-0.25}^{+0.26}$ & $10.94_{-3.10}^{+3.28}$ & $1.02(68)$ & $-0.24_{-1.10}^{+1.64}$ & $-2.55_{-\infty}^{+0.29}$ & $35.41_{-8.53}^{+18.41}$ & $0.98(67)$ \\
3 & $10.98_{-2.14}^{+2.56}$ & $0.74_{-0.27}^{+0.31}$ & $2.16_{-0.25}^{+0.40}$ & $32.99_{-20.23}^{+65.24}$ & $0.94(69)$ & $9.07_{-1.32}^{+1.39}$ & $0.80_{-0.24}^{+0.26}$ & $9.21_{-2.87}^{+3.02}$ & $0.99(70)$ & $-0.85_{-0.54}^{+0.80}$ & $-3.07_{-\infty}^{+0.62}$ & $39.74_{-7.77}^{+8.60}$ & $0.93(69)$ \\
4 & $6.82_{-1.13}^{+1.69}$ & $0.61_{-0.26}^{+0.28}$ & $1.85_{-0.34}^{+0.27}$ & $9.05_{-7.42}^{+17.99}$ & $0.92(134)$ & $6.76_{-1.08}^{+1.17}$ & $0.63_{-0.20}^{+0.21}$ & $8.13_{-2.71}^{+2.89}$ & $0.91(135)$ & $0.52_{-1.60}^{+3.16}$ & $-2.47_{-0.54}^{+0.25}$ & $24.48_{-5.39}^{+9.05}$ & $0.92(134)$ \\
5 & $11.36_{-1.74}^{+2.01}$ & $1.12_{-0.29}^{+0.30}$ & $2.08_{-0.13}^{+0.16}$ & $49.09_{-19.27}^{+31.10}$ & $0.91(147)$ & $9.18_{-1.02}^{+1.06}$ & $1.19_{-0.26}^{+0.27}$ & $18.14_{-3.07}^{+3.19}$ & $0.97(148)$ & $-1.06_{-0.32}^{+0.46}$ & $-2.92_{-\infty}^{+0.45}$ & $43.19_{-7.51}^{+7.34}$ & $0.88(147)$ \\
6 & $22.61_{-0.99}^{+1.00}$ & $8.52_{-0.65}^{+0.68}$ & $1.77_{-0.06}^{+0.07}$ & $35.88_{-7.30}^{+9.67}$ & $1.56(178)$ & $23.10_{-0.78}^{+0.82}$ & $8.92_{-0.49}^{+0.50}$ & $43.47_{-3.33}^{+3.37}$ & $1.56(179)$ & $-0.45_{-0.09}^{+0.09}$ & $-10.0$ & $102.66_{-4.34}^{+4.77}$ & $1.16(178)$ \\
7 & $23.69_{-0.73}^{+0.74}$ & $13.67_{-0.74}^{+0.77}$ & $1.73_{-0.05}^{+0.06}$ & $41.65_{-6.89}^{+8.52}$ & $1.86(182)$ & $24.34_{-0.59}^{+0.61}$ & $14.53_{-0.58}^{+0.58}$ & $55.80_{-3.53}^{+3.57}$ & $1.89(183)$ & $-0.31_{-0.08}^{+0.08}$ & $-3.82_{-\infty}^{+0.52}$ & $105.94_{-3.95}^{+4.20}$ & $1.28(182)$ \\
8 & $19.77_{-0.82}^{+0.84}$ & $8.93_{-0.58}^{+0.60}$ & $1.76_{-0.04}^{+0.05}$ & $53.13_{-7.89}^{+9.22}$ & $1.87(180)$ & $20.51_{-0.65}^{+0.67}$ & $9.46_{-0.49}^{+0.50}$ & $66.35_{-3.90}^{+3.95}$ & $1.90(181)$ & $-0.61_{-0.08}^{+0.09}$ & $-3.30_{-1.09}^{+0.41}$ & $91.76_{-4.95}^{+4.85}$ & $1.24(180)$ \\
9 & $14.41_{-0.77}^{+0.81}$ & $5.93_{-0.44}^{+0.44}$ & $1.86_{-0.04}^{+0.05}$ & $75.42_{-12.23}^{+14.09}$ & $1.69(175)$ & $14.06_{-0.54}^{+0.57}$ & $5.87_{-0.42}^{+0.43}$ & $67.70_{-4.14}^{+4.20}$ & $1.69(176)$ & $-0.88_{-0.08}^{+0.14}$ & $-9.37_{-\infty}^{+19.37}$ & $70.82_{-2.87}^{+3.00}$ & $1.09(175)$ \\
10 & $12.64_{-1.05}^{+1.17}$ & $3.36_{-0.37}^{+0.38}$ & $1.84_{-0.06}^{+0.06}$ & $54.40_{-11.52}^{+13.64}$ & $1.58(164)$ & $12.41_{-0.72}^{+0.78}$ & $3.35_{-0.37}^{+0.37}$ & $51.03_{-3.86}^{+3.92}$ & $1.57(165)$ & $-1.02_{-0.10}^{+0.11}$ & $-9.37_{-\infty}^{+19.37}$ & $66.62_{-3.22}^{+4.25}$ & $1.23(164)$ \\
11 & $11.33_{-1.06}^{+1.19}$ & $1.97_{-0.30}^{+0.31}$ & $2.01_{-0.12}^{+0.14}$ & $40.27_{-15.24}^{+22.81}$ & $1.32(149)$ & $10.27_{-0.70}^{+0.73}$ & $2.03_{-0.29}^{+0.30}$ & $19.73_{-3.23}^{+3.33}$ & $1.36(150)$ & $-0.61_{-0.31}^{+0.45}$ & $-3.11_{-\infty}^{+0.44}$ & $45.39_{-5.74}^{+5.17}$ & $1.26(149)$ \\
12 & $10.03_{-1.07}^{+1.19}$ & $1.42_{-0.27}^{+0.29}$ & $2.21_{-0.24}^{+0.36}$ & $37.04_{-21.85}^{+59.08}$ & $0.99(141)$ & $8.96_{-0.72}^{+0.74}$ & $1.50_{-0.25}^{+0.26}$ & $8.31_{-2.79}^{+2.91}$ & $1.03(142)$ & $-0.51_{-0.37}^{+0.43}$ & $-10.0$ & $39.19_{-3.41}^{+3.82}$ & $0.95(141)$ \\
13 & $7.61_{-1.11}^{+1.35}$ & $0.65_{-0.19}^{+0.17}$ & $3.65_{-1.58}^{+3.10}$ & $164.92$ & $1.13(127)$ & $7.29_{-0.98}^{+1.02}$ & $0.65_{-0.20}^{+0.20}$ & $0.44_{-0.44}^{+2.39}$ & $1.15(128)$ & $1.18_{-1.59}^{+1.72}$ & $-4.48_{-\infty}^{+1.25}$ & $27.93_{-3.58}^{+4.48}$ & $1.14(127)$ \\
\hline
\end{tabular}
\end{scriptsize}
\label{081221_a}
 
\end{sidewaystable}

\begin{table}\centering
 \caption{$\chi_{red}^2$ of different models for time-resolved spectral analysis of GRB 081221}
 \begin{tabular}{c|c|c|c|c}
\hline
 Method & \multicolumn{2}{c|}{3 Second time bins} & \multicolumn{2}{c}{1 Second time bins} \\
\cline{2-5}

       & $\langle \chi_{red}^2 \rangle$ of full GRB &  $\langle \chi_{red}^2 \rangle$ (2nd pulse)  & $\langle \chi_{red}^2 \rangle$ of full GRB &  $\langle \chi_{red}^2 \rangle$ (2nd pulse)\\
\hline
\hline 
BBPL ($\Gamma$ free) & $1.31 \pm 0.35$ & $1.52 \pm 0.32$ & $1.11 \pm 0.26$ & $1.21 \pm 0.28$\\
BBPL ($\Gamma$ frozen) & $1.30 \pm 0.35$ & $1.50 \pm 0.33$ & $1.16 \pm 0.27$ & $1.22 \pm 0.29$\\
Band & $1.09 \pm 0.14$ & $1.17 \pm 0.11$ & $1.00 \pm 0.16$ & $1.04 \pm 0.18$ \\
mBBPL & $1.15 \pm 0.14$ & $1.23 \pm 0.13$ & $1.07 \pm 0.17$ & $1.06 \pm 0.19$ \\
2BBPL & $1.09 \pm 0.15$ & $1.17 \pm 0.13$ & $1.02 \pm 0.16$ & $1.05 \pm 0.17$ \\
\hline
\end{tabular}
\label{av_chi}
\end{table}

\clearpage
\thispagestyle{empty}

\begin{sidewaystable}
\begin{small}
\caption{Study of spectral evolution in pulse 2 (17.0 to 40.55 s) of GRB 081221 (neglecting 30-40 keV)
}
\hspace{-2.0cm}
\begin{tabular}{ccccccccccc}

\hline

\hline
Model & $\chi^2$ (dof) & $\mu$ & $\nu$ & $\alpha$ & $\beta$ & $E_{peak}$ $^a$ & p & $\Gamma$ & $kT_h $/$kT_{in} $/$kT$ $ ^a$  & $kT_l$ $^a$\\
\hline
\hline
\multicolumn{9}{l}{Case I: Count per time bin $\gtrsim$ 3000 --- Rising part (17.0 to 21.45 s 3 bins)}\\
\hline

Band & 364.41 (354) & $1.0 \pm0.3$ & --- & $-0.44_{-0.03}^{+0.06}$ & $-7.61_{-\infty}^{+2.14}$ & $98.59_{-3.02}^{+2.75}$ & --- & --- & --- & --- \\
BBPL & 455.92 (354) & $0.5 \pm0.3$ & $3.2 \pm1.0$ & ---      & ---  &   --- & --- &  $1.89_{-0.04}^{+0.05}$ &   $24.16_{-0.63}^{+0.64}$ & --- \\
mBBPL & 355.68 (353) & $0.9 \pm 0.2$ & $0.8 \pm 1.0$ & --- & --- & --- &  $0.81_{-0.04}^{+0.08}$ & $1.81_{-0.11}^{+0.20}$ & $40.02_{-2.04}^{+2.87}$ & --- \\
2BBPL & 351.78 (352) & $0.6 \pm0.1$ & $4.7\pm 1.0$ & --- & --- & --- & --- & $1.94_{-0.11}^{+0.16}$ & $28.73_{-1.44}^{+1.67}$ & $9.75_{-1.03}^{+1.14}$ \\

\hline
\multicolumn{9}{l}{Case I: Count per time bin $\gtrsim$ 3000 --- Falling part (21.55 to 40.55 s 9 bins)}\\
\hline
Band & 1188.17 (993) & $-2.1 \pm0.1$ & --- & $-0.68 \pm 0.05$ & $-3.55_{-0.44}^{+0.26}$ & $115.5_{-3.2}^{+3.0}$ & --- & --- & --- & --- \\
BBPL & 1557.25 (993) & $-1.9 \pm0.1$ & $-3.2 \pm0.4$ & ---      & ---  &   --- & --- &  $2.02_{-0.02}^{+0.03}$ &   $26.81_{-0.57}^{+0.58}$ & --- \\
mBBPL & 1223.39 (992) & $-2.0 \pm 0.2$ & $3.5 \pm 0.3$ & --- & --- & --- &  $0.74_{-0.03}^{+0.02}$ & $2.03_{-0.06}^{+0.09}$ & $49.93_{-1.44}^{+2.82}$ & --- \\
2BBPL & 1147.53 (991) & $-1.9 \pm0.1$ & $-3.1\pm 0.4$ & --- & --- & --- & --- & $2.15_{-0.08}^{+0.09}$ & $38.13_{-1.52}^{+1.63}$ & $13.33_{-0.68}^{+0.71}$ \\

\hline
\multicolumn{9}{l}{Case II: Count per time bin $\gtrsim$ 1000 --- Rising part (17.0 to 21.45 s 10 bins)}\\

\hline
Band & 1247.91 (1187) & $1.5 \pm0.3$ & --- & $-0.44_{-0.06}^{+0.05}$ & $-9.15_{-\infty}^{+4.02}$ & $90.35_{-2.48}^{+2.96}$ & --- & --- & --- & --- \\
BBPL & 1328.50 (1187) & $1.0 \pm0.3$ & $4.4 \pm 0.8$ & ---      & ---  &   --- & --- &  $1.89_{-0.04}^{+0.05}$ &   $22.55_{-0.63}^{+0.64}$ & --- \\
mBBPL & 1239.20 (1186) & $0.9 \pm 0.1$ & $1.3 \pm 0.5$ & --- & --- & --- &  $0.87\pm0.09$ & $1.96_{-0.14}^{+0.28}$ & $39.01_{-1.92}^{+3.48}$ & --- \\
2BBPL & 1222.76 (1185) & $0.4 \pm0.6$ & $9.0\pm 2.0$ & --- & --- & --- & --- & $2.07_{-0.16}^{+0.37}$ & $30.30_{-1.54}^{+1.73}$ & $9.96_{-0.96}^{+1.01}$ \\
\hline

\multicolumn{9}{l}{Case II: Count per time bin $\gtrsim$ 1000 --- Falling part (21.55 to 40.55 s 29 bins)}\\

\hline
Band & 3743.36 (3448) & $-2.5 \pm0.1$ & --- & $-0.75_{-0.05}^{+0.06}$ & $-3.56_{-0.77}^{+0.31}$ & $125.7\pm3.9$ & --- & --- & --- & --- \\
BBPL & 4133.63 (3448) & $-2.1 \pm0.1$ & $-3.6 \pm0.2$ & ---      & ---  &   --- & --- &  $2.04\pm0.03$ &   $28.15_{-0.67}^{+0.68}$ & --- \\
mBBPL & 3804.05 (3447) & $-2.0 \pm 0.1$ & $3.0 \pm 0.5$ & --- & --- & --- &  $0.72_{-0.02}^{+0.03}$ & $2.22_{-0.08}^{+0.11}$ & $53.00_{-2.39}^{+2.13}$ & --- \\
2BBPL & 3690.21 (3446) & $-2.0 \pm0.2$ & $-3.6\pm 0.4$ & --- & --- & --- & --- & $2.31_{-0.11}^{+0.13}$ & $41.40_{-1.61}^{+1.69}$ & $13.72_{-0.62}^{+0.65}$ \\
\hline
\end{tabular}
\label{p2}
 
\begin{footnotesize} $^a$ The values quoted are for the first time bin. \\

\end{footnotesize}
\end{small}
\end{sidewaystable}

\clearpage
\thispagestyle{empty}

\begin{sidewaystable}
\begin{small}

\caption{Study of spectral evolution in pulse 1 (-1.0 to 12.05 s) of GRB 081221. The bins are obtained by requiring $\gtrsim$ 1000 counts per bin
(30-40 keV bins are neglected)}
\begin{tabular}{ccccccccccc}

\hline

\hline
Model & $\chi^2$ (dof) & $\mu$ & $\nu$ & $\alpha$ & $\beta$ & $E_{peak}$ $^a$ & p & $\Gamma$ & $kT_h $/$kT_{in} $/$kT$ $ ^a$  & $kT_l$ $^a$\\
\hline
\hline
\multicolumn{9}{l}{Rising part (-1 to 2.15 s 1 bin)}\\
\hline

Band & 115.78 (116) & --- & --- & $-0.55_{-0.22}^{+0.26}$ & $-10.0$ & $170.3_{-22.7}^{+30.7}$ & --- & --- & --- & --- \\
BBPL & 109.67 (116) & --- & --- & ---      & ---  &   --- & --- &  $1.93_{-0.21}^{+0.35}$ &   $38.27_{-3.76}^{+4.08}$ & --- \\
mBBPL & 110.27 (115) & --- & --- & --- & --- & --- &  $0.98_{-0.28}^{+\infty}$ & $2.15_{-4.46}^{+\infty}$ & $62.78_{-7.22}^{+18.14}$ & --- \\
2BBPL & 103.05 (114) & --- & --- & --- & --- & --- & --- & $1.74_{-3.04}^{+\infty}$ & $38.47_{-3.66}^{+4.40}$ & $6.57_{-1.64}^{+3.17}$ \\

\hline
\multicolumn{9}{l}{Falling part (2.25 to 12.05 s 4 bins)}\\
\hline
Band & 544.65 (473) & $-0.7 \pm0.1$ & --- & $-0.86_{-0.19}^{+0.22}$ & $-3.61_{-\infty}^{+0.69}$ & $82.02_{-7.43}^{+7.53}$ & --- & --- & --- & --- \\
BBPL & 571.47 (473) & $-0.7 \pm0.1$ & $-0.2 \pm0.4$ & ---      & ---  &   --- & --- &  $2.09_{-0.11}^{+0.13}$ &   $19.62_{-1.77}^{+1.88}$ & --- \\
mBBPL & 548.22 (472) & $-0.7 \pm 0.2$ & $2.5 \pm 0.8$ & --- & --- & --- &  $0.63_{-0.03}^{+0.10}$ & $1.51_{-\infty}^{+0.52}$ & $41.63_{-6.76}^{+6.49}$ & --- \\
2BBPL & 544.79 (471) & $-0.7 \pm0.2$ & $-0.1\pm 0.5$ & --- & --- & --- & --- & $2.04_{-0.14}^{+0.28}$ & $28.59_{-4.45}^{+5.99}$ & $9.95_{-1.89}^{+2.30}$ \\
\hline
\end{tabular}
\label{p1}
 
\begin{footnotesize}

 $^a$ The values quoted are for the first time bin. \\

\end{footnotesize}
\end{small}

\end{sidewaystable}

\begin{table}\centering
 \caption{Comparison of different model fits at different episodes of GRB 081221}
 \begin{tabular}{c|c|c|c|c}
\hline
Region & Model$_2$/Model$_1$ & p & $\sigma$ & C.L.\\
\hline
\hline 
Pulse 1, Rising part & BBPL/Band & 0.385 & 0.87 & 61.46\% \\
\cline{2-5}
 & mBBPL/Band & 0.415 & 0.81 & 58.50\% \\
\cline{2-5}
 & 2BBPL/Band & 0.298 & 1.04 & 70.17\% \\
\cline{2-5}
 & 2BBPL/BBPL & 0.029 & 2.19 & 97.12\% \\
\hline

Pulse 1, Falling part & Band/BBPL & 0.301 & 1.03 & 69.93\% \\
\cline{2-5}
 & mBBPL/BBPL & 0.334 & 0.965 & 66.57\% \\
\cline{2-5}
 & 2BBPL/BBPL & 1.29$\times10^{-5}$ & 4.36 & 99.99\% \\
\cline{2-5}
 & Band/2BBPL & 0.480 & 0.705 & 51.95\% \\
\hline

Pulse 2, Rising part & Band/BBPL & 0.018 & 2.37 & 98.23\% \\
\cline{2-5}
 ($\gtrsim$ 3000 counts/bin) & mBBPL/BBPL & 0.011 & 2.55 & 98.93\% \\
\cline{2-5}
 & 2BBPL/BBPL & 1.5$\times10^{-20}$ & 9.29 & 100\% \\
\cline{2-5}
 & 2BBPL/Band & 0.390 & 0.86 & 60.94\% \\
\hline

Pulse 2, Falling part & Band/BBPL & 1.04$\times10^{-5}$ & 4.41 & 99.99\% \\
\cline{2-5}
 ($\gtrsim$ 3000 counts/bin) & mBBPL/BBPL & 7.86$\times10^{-5}$ & 3.95 & 99.99\% \\
\cline{2-5}
 & 2BBPL/BBPL & 1.99$\times10^{-66}$ & 17.22 & 100\% \\
\cline{2-5}
 & 2BBPL/Band & 0.303 & 1.03 & 69.71\% \\
\hline

Pulse 2, Rising part & Band/BBPL & 0.172 & 1.36 & 82.76\% \\
\cline{2-5}
 ($\gtrsim$ 1000 counts/bin) & mBBPL/BBPL & 0.118 & 1.56 & 88.17\% \\
\cline{2-5}
 & 2BBPL/BBPL & 4.55$\times10^{-22}$ & 9.66 & 100\% \\
\cline{2-5}
 & 2BBPL/Band & 0.374 & 0.89 & 62.60\% \\
\hline

Pulse 2, Falling part & Band/BBPL & 0.0018 & 3.12 & 99.82\% \\
\cline{2-5}
 ($\gtrsim$ 1000 counts/bin) & mBBPL/BBPL & 0.0075 & 2.67 & 99.24\% \\
\cline{2-5}
 & 2BBPL/BBPL & 1.23$\times10^{-85}$ & 19.61 & 100\% \\
\cline{2-5}
 & 2BBPL/Band & 0.343 & 0.95 & 65.64\% \\
\hline
\end{tabular}
\label{sigma_level}
\end{table}

\clearpage
\thispagestyle{empty}

\begin{sidewaystable}
\begin{small}

\caption{Study of spectral evolution in pulse 1 (-1.0 to 40.85 s) of GRB 090618. The bins are obtained by requiring $\gtrsim$ 1000 counts 
per bin
}
\begin{tabular}{ccccccccccc}

\hline

\hline
Model & $\chi^2$ (dof) & $\mu$ & $\nu$ & $\alpha$ & $\beta$ & $E_{peak}$ $^a$ & p & $\Gamma$ & $kT_h $/$kT_{in} $/$kT$ $ ^a$  & $kT_l$ $^a$\\
\hline
\hline
\multicolumn{9}{l}{Rising part (-1 to 14.15 s 10 bins)}\\
\hline

Band & 623.94 (547) & $-0.8 \pm0.2$ & --- & $-0.46_{-0.09}^{+0.10}$ & $-3.07_{-0.61}^{+0.32}$ & $344.9_{-22.0}^{+22.1}$ & --- & --- & --- & --- \\
BBPL & 661.77 (547) & $-0.6 \pm0.2$ & $-1.4 \pm0.4$ & ---      & ---  &   --- & --- &  $1.71_{-0.04}^{+0.05}$ &   $64.79_{-2.46}^{+2.50}$ & --- \\
mBBPL & 621.93 (546) & $-0.7 \pm0.2$ & $-0.5 \pm2.0$ & --- & --- & --- &  $0.83_{-0.06}^{+0.09}$ & $1.69_{-0.10}^{+0.19}$ & $125.1_{-8.7}^{+12.5}$ & --- \\
2BBPL & 624.55 (545) & $-0.7 \pm0.2$ & $-1.4 \pm0.8$ & --- & --- & --- & --- & $1.72_{-0.08}^{+0.11}$ & $79.69_{-5.39}^{+9.19}$ & $25.88_{-5.08}^{+9.45}$ \\

\hline
\multicolumn{9}{l}{Falling part (14.15 to 40.85 s 11 bins)}\\
\hline
Band & 602.53 (571) & $-1.0 \pm0.3$ & --- & $-0.79_{-0.11}^{+0.13}$ & $-3.02_{-1.10}^{+0.38}$ & $186.2_{-19.2}^{+18.3}$ & --- & --- & --- & --- \\
BBPL & 642.37 (571) & $-0.9 \pm0.3$ & $-1.4 \pm0.7$ & ---      & ---  &   --- & --- &  $1.83 \pm 0.05$ &   $37.95_{-2.50}^{+2.67}$ & --- \\
mBBPL & 602.30 (570) & $-1.1 \pm 0.2$ & $1.5 \pm 1.5$ & --- & --- & --- &  $0.69_{-0.03}^{+0.06}$ & $1.70_{-0.19}^{+0.21}$ & $85.81_{-7.67}^{+12.27}$ & --- \\
2BBPL & 593.74 (569) & $-0.3 \pm0.3$ & $-2.9\pm 1.0$ & --- & --- & --- & --- & $2.11_{-0.12}^{+0.10}$ & $50.88_{-4.66}^{+4.88}$ & $15.33_{-1.65}^{+1.93}$ \\
\hline
\end{tabular}
\label{090618_a}
 
\begin{footnotesize}

 $^a$ The values quoted are for the first time bin. \\

\end{footnotesize}
\end{small}

\end{sidewaystable}

\clearpage
\thispagestyle{empty}

\begin{sidewaystable}
\begin{small}
\caption{Study of spectral evolution in pulse 2 (61 to 75.0 s) of GRB 090618. The bins are obtained by requiring $\gtrsim$ 2000 counts 
per bin
}
\hspace{-2.0cm}
\begin{tabular}{ccccccccccc}

\hline

\hline
Model & $\chi^2$ (dof) & $\mu$ & $\nu$ & $\alpha$ & $\beta$ & $E_{peak}$ $^a$ & p & $\Gamma$ & $kT_h $/$kT_{in} $/$kT$ $ ^a$  & $kT_l$ $^a$\\
\hline
\hline
\multicolumn{9}{l}{Rising part (61 to 64.35 s 1 bin)}\\
\hline

Band & 2012.93 (1707) & $8.0 \pm3.0$ & --- & $-0.68_{-0.05}^{+0.06}$ & $-2.49_{-0.12}^{+0.09}$ & $192.5_{-10.3}^{+11.3}$ & --- & --- & --- & --- \\
BBPL & 2173.47 (1707) & $5.0 \pm1.0$ & $6.5 \pm5.0$ & ---      & ---  &   --- & --- &  $1.64 \pm 0.02$ &   $42.82_{-1.16}^{+1.17}$ & --- \\
mBBPL & 2099.21 (1706) & $8.0 \pm1.5$ & $10.0 \pm5.0$ & --- & --- & --- &  $0.68 \pm 0.02$ & $1.29 \pm 0.08$ & $97.1_{-4.9}^{+6.4}$ & --- \\
2BBPL & 2030.83 (1705) & $6.0 \pm1.5$ & $3.0 \pm3.0$ & --- & --- & --- & --- & $1.78_{-0.04}^{+0.05}$ & $115.56_{-12.09}^{+13.33}$ & $33.19_{-1.73}^{+1.71}$ \\

\hline
\multicolumn{9}{l}{Falling part (64.35 to 74.95 s 24 bins)}\\
\hline
Band & 1574.79 (1317) & $-14.0 \pm 1.0$ & --- & $-0.88 \pm 0.03$ & $-2.74_{-0.10}^{+0.08}$ & $262.8_{-7.9}^{+8.5}$ & --- & --- & --- & --- \\
BBPL & 2402.54 (1317) & $-5.3 \pm0.3$ & $-6.9 \pm0.9$ & ---      & ---  &   --- & --- &  $1.78 \pm 0.01$ &   $53.91_{-1.00}^{+1.01}$ & --- \\
mBBPL & 1794.05 (1316) & $-8.0 \pm 3.0$ & $15.0 \pm 5.0$ & --- & --- & --- &  $0.65_{-0.005}^{+0.006}$ & $1.58_{-0.10}^{+0.03}$ & $157.36_{-5.68}^{+4.18}$ & --- \\
2BBPL & 1794.06 (1315) & $-5.5 \pm0.5$ & $-4.0\pm 0.5$ & --- & --- & --- & --- & $1.78 \pm 0.02$ & $73.08_{-2.57}^{+3.04}$ & $21.72_{-1.44}^{+1.68}$ \\
\hline
\end{tabular}
\label{090618_b}
 
\begin{footnotesize}

 $^a$ The values quoted are for the first time bin. \\

\end{footnotesize}
\end{small}

\end{sidewaystable}

\clearpage

\clearpage

\clearpage

\clearpage

\clearpage

\clearpage

\clearpage

\clearpage

\clearpage

\clearpage

\end{document}